\documentclass[12pt]{article}

\pdfoutput=1
\usepackage{graphicx}
\usepackage{amssymb}
\usepackage{amsmath}
\usepackage{color}
\usepackage{subfigure}
\usepackage[colorlinks=true,linkcolor=blue,citecolor=blue]{hyperref}

\begin{document}

\newcommand{\be}{\begin{equation}}
\newcommand{\ee}{\end{equation}}


\begin{titlepage}

\begin{flushright}
ICRR-Report-603-2011-20  \\
IPMU 11-0205\\
UT-11-43
\end{flushright}

\vskip 2cm

\begin{center}

{\large \bf
Revisiting the cosmological coherent oscillation
}

\vspace{1cm}

{Masahiro Kawasaki$^{(a, b)}$, Naoya Kitajima$^{(a)}$
and Kazunori Nakayama$^{(c,b)}$}

\vskip 1.0cm

{\it
$^a$Institute for Cosmic Ray Research,
     University of Tokyo, Kashiwa, Chiba 277-8582, Japan\\
$^b$Kavli Institute for the Physics and Mathematics of the Universe, 
     University of Tokyo, Kashiwa, Chiba 277-8568, Japan\\
$^c$Department of Physics, University of Tokyo, Bunkyo-ku, Tokyo 113-0033, Japan
}

\vskip 1.0cm

\begin{abstract}
It is often the case that scalar fields are produced in the early Universe in the form of coherent oscillation.
These scalar fields may have huge abundances and affect the evolution of the Universe.
In particular, if the lifetime is long enough, they may cause cosmological disasters.
We revisit the issue of coherent oscillation of the scalar field when it couples with another oscillating scalar field, and find a situation that the abundance, or the amplitude of the oscillation, is significantly reduced by a variant type of the adiabatic suppression mechanism.
As a concrete example, it is applied to the saxion, a flat direction in the supersymmetric axion model, and we show that the cosmological saxion problem is solved in a particular setup.
\end{abstract}

\end{center}

\end{titlepage}

\newpage
\tableofcontents

\vspace{1cm}

\section{Introduction} \label{intro}

In the theory beyond the standard model such as supersymmetry (SUSY), supergravity or string theory, various scalar fields appear, which are often called moduli, and they may give significant effects on the cosmological scenario~\cite{Coughlan:1983ci}.
Moduli are in general stabilized at the high energy minimum in the early Universe, which are displaced from the low energy minimum.
At some epoch, the moduli start the coherent oscillation around the true minimum with large initial amplitude and hence the Universe may be dominated by the moduli.
Some of them may acquire the mass only from the SUSY breaking effect and it is of the order of the gravitino mass~\cite{Banks:1993en}.
In addition, they may have interactions with the standard model particles typically suppressed by the Planck scale or some high energy scales.
Hence their lifetimes are long enough and their decays may significantly affect the cosmology.
The abundance of late-decaying moduli is strongly constrained by observations such as 
the light element abundances synthesized through big bang nucleosynthesis (BBN), 
the distortion of the cosmic microwave background (CMB) spectrum, 
or diffuse X($\gamma$) ray background~\cite{Ellis:1990nb,Asaka:1999xd}.
If the modulus lifetime is longer than the present age of the Universe, it contributes to the cold dark matter (CDM) density, and its abundance is constrained so as not to exceed the present matter density.
 
Aside from the moduli, there may be many singlet scalar fields which more or less have similar properties to the moduli.
We focus on such a scalar field $\phi$ whose lifetime is  so long that its decay might affect the cosmology after BBN.
Generally, the energy density to entropy ratio of $\phi$ is calculated as
\be
	\frac{\rho_\phi}{s} = 
	\frac{1}{8} T_R \bigg( \frac{\phi_i}{M_P} \bigg)^2 
	\simeq 1 \times 10^4\,\mathrm{GeV} 
	\bigg( \frac{T_R}{10^5\,\mathrm{GeV}} \bigg) 
	\bigg( \frac{\phi_i}{M_P} \bigg)^2,   \label{abundance_nodamp1}
\ee
for $m_\phi > \Gamma_I$, where $m_\phi$ is the mass of $\phi$, $\Gamma_I$ is the inflaton decay rate $T_R$ is the reheating temperature and $\phi_i$ is the initial amplitude of $\phi$.
For $m_\phi < \Gamma_I$, we have
\be
	\frac{\rho_\phi}{s} = \frac{1}{8} T_\mathrm{osc} 
	\bigg( \frac{\phi_i}{M_P} \bigg)^2
	\simeq 1 \times 10^{9}\,\mathrm{GeV} 
	\bigg( \frac{g_*}{100} \bigg)^{-1/4} 
	\bigg( \frac{m_\phi}{100\,\mathrm{GeV}} \bigg)^{1/2}
	\bigg( \frac{\phi_i}{M_P} \bigg)^2,
	\label{abundance_nodamp2}
\ee
where $T_\mathrm{osc}$ represents the temperature when $\phi$ starts to oscillate.
Comparing the above estimates to the observational constraints, $\phi$ causes cosmological difficulties unless its abundance is somehow suppressed.
A late-time entropy production process may act as an efficient dilution mechanism for such coherent oscillations, but it simultaneously dilutes the preexisting baryon asymmetry.
Although some elaborate baryogenesis mechanisms may work even under the entropy production, we here consider another possibility.

In this respect, Linde~\cite{Linde:1996cx} proposed an interesting scenario that the modulus abundance is significantly suppressed without any entropy production process.
The idea is that if the modulus obtains a mass squared of $\sim c^2 H^2$ where $H$ is the Hubble parameter and $c\gtrsim \mathcal O(10)$, the modulus adiabatically follows the temporal minimum of the potential toward the true minimum.
As a result, the amplitude of the coherent oscillation is exponentially suppressed.
This mechanism has been recently investigated in detail in Ref.~\cite{Nakayama:2011wqa} where it was shown that the modulus oscillation is generically induced at the end of inflation to a small but non-negligible amount.
Still, the adiabatic suppression mechanism works as an efficient way to significantly reduce the modulus abundance
for solving the cosmological moduli problem~\cite{arXiv:1112.0418}.

The essential point for the adiabatic suppression is that the scalar obtains a effective mass larger than the Hubble parameter, which gradually disappears after the Hubble friction becomes inefficient.
Thus thermal mass for the scalar may also cause an adiabatic suppression~\cite{Nakayama:2008ks}.
In this paper, we propose a variant method to suppress the amplitude of the coherent oscillation of the scalar field, although it may not be the moduli in an original sense.

Let us consider a scalar field $\phi$, whose abundance we want to know, and suppose that its true minimum is displaced from the initial position determined during/after inflation.
In this setup, it is not hard to imagine that the coherent oscillation of $\phi$ is induced when the Hubble parameter becomes equal to the mass of $\phi$ and the resulting abundance is estimated by 
(\ref{abundance_nodamp1}) and (\ref{abundance_nodamp2}).
The situation significantly changes if the scalar field $\phi$ couples to another scalar field $S$, which has a large field value and is oscillating.
The coupling induces an effective mass of $\phi$, and it can be large enough to overcome the Hubble parameter.
Hence the scalar field $\phi$ follows the time-dependent minimum of the potential adiabatically.
As a result, the amplitude of the coherent oscillation of $\phi$ is continuously damped by the expansion of the Universe and the energy density of $\phi$ is highly suppressed compared with the one without the coupling.
This phenomenon was already noted in Ref.~\cite{Kawasaki:2010gv} in a context of SUSY axion model. Here we wish to generalize and reformulate the arguments there.

This paper is organized as follows.
In Sec.~\ref{toy_model}, our proposal to suppress the scalar field abundance is discussed by using a simple toy model.
Essential ingredients are all contained in this simple toy model.
An application of our mechanism to the SUSY axion model is discussed in Sec.~\ref{axion}.
Sec.~\ref{conc} is devoted to conclusions.

\section{Adiabatic suppression mechanism in a simple toy model} \label{toy_model}

In this section, we illustrate the mechanism to suppress the oscillation amplitude of the scalar field.
The essence of the idea is that if the mass of the modulus field is much larger than the Hubble parameter, the field oscillation follows its time-dependent minimum of the potential adiabatically.
This is because the potential around the temporal minimum is steep enough to overcome the Hubble friction. 
Then the resultant amplitude of the oscillation becomes very small compared with the general case considered in the previous section.  
Here we introduce an another scalar field, which supplies the large mass to the scalar field.
To see the suppression of the amplitude, we follow the scalar field dynamics within a simple toy model.

\subsection{Toy model}

Let us consider a simple toy model, where two gauge singlet real scalar fields are introduced.
One is the field $\phi$ whose abundance is what we focus on, and the other is the heavy field $S$ which supplies a large mass to $\phi$.
The scalar potential is given by
\be
	V = \frac{1}{2} m_\phi^2 (\phi - \phi_0)^2 + \frac{1}{2} m_S^2 S^2 
	+ \frac{1}{2} \lambda^2 S^2 \phi^2,   \label{toy_pot}
\ee
where $m_\phi$ and $m_S$ are the masses of $\phi$ and $S$, respectively and $\lambda$ is a coupling constant which is assumed to be positive for simplicity.
This is the most general form respecting the $Z_2$ symmetry under which $S$ transforms as $S\to -S$, up to the quadratic terms in $\phi$ and $S$.
The true minimum of the potential lies at $S=0$ and $\phi=\phi_0$.
In the early Universe, however, both $\phi$ and $S$ may be displaced from the minimum.
In particular, if $S$ has a large amplitude, the temporal minimum of $\phi$ is calculated as 
\be
	\langle \phi \rangle = \frac{m_\phi^2}{m_\phi^2 + \lambda^2 S^2} \phi_0.
	\label{phimin}
\ee
This is time-dependent and estimated as $\langle \phi \rangle \simeq 0$ for $|\lambda S| \gg m_\phi$ and $\langle \phi \rangle \simeq \phi_0$ for $m_\phi \gg |\lambda S|$.
The equations of motion are given by
\be
	\ddot\phi + 3H\dot\phi + m_\phi^2 ( \phi - \phi_0) +
	 \lambda^2 S^2 \phi = 0,   \label{EOM_phi}
\ee
\be
	\ddot{S} + 3H \dot{S} + (m_S^2 + \lambda^2 \phi^2) S = 0,   \label{EOM_S}
\ee
where $H$ is the Hubble parameter.
In order to give the large effective mass to $\phi$ enough to overcome the Hubble parameter, we assume that $S$ is initially far displaced from the origin.
Then the effective mass of $\phi$ is given by $m_{\phi}^\mathrm{eff} \simeq \lambda S$ for $|\lambda S |\gg m_\phi$.
Hence, the required condition for $\phi$ to follow the time-dependent minimum adiabatically is $|\lambda S |\gg H$.
Under these assumptions, we follow the field dynamics by solving Eqs.~(\ref{EOM_phi}) and (\ref{EOM_S}) and calculate the resultant abundance.

\subsection{The scalar field dynamics}

We investigate the dynamics of the scalar fields after the field $S$ starts to oscillate.
For simplicity, we assume that the Universe is inflaton oscillation-dominated in the era we are interested in and the reheating takes place well after that.\footnote{
Both $\phi$ and $S$ are not assumed to be the inflaton.
}
Because $\phi$ is initially settled on the origin, the mass of $S$ is given by $m_S$, and hence the $S$ begins to oscillate at $H=m_S$.
The solution of the equation of motion (\ref{EOM_S}) after $S$ starts to oscillate is
\be
	S=A_S \cos(m_S t) ~~\text{with}~~A_S = S_i \frac{H}{m_S} ,   \label{sol_S}
\ee
for $H<m_S$ where $S_i$ represent the initial amplitude of $S$, which we regard as a free parameter.
After $S$ starts to oscillate, its amplitude decreases in proportional to $H$.
Therefore, the effective mass of the $\phi$, $m_\phi^{\rm eff}\sim \lambda A_S$, always exceed the Hubble parameter if the following condition is satisfied : $\lambda S_i > m_S$.
Then, based on the similar arguments performed in Refs.~\cite{Linde:1996cx,Nakayama:2011wqa}, the $\phi$ abundance is significantly suppressed by the adiabatic suppression mechanism, if the $S$ decay rate $(\Gamma_S)$ is so small that $S$ decays after the $\phi$ relaxes to the minimum.
This condition is rewritten as $\Gamma_S < H_{\rm adi}\equiv m_Sm_\phi/(\lambda S_i)$.
If $S$ has a large mass and decays well before BBN, it does not cause significant cosmological effects.
Then cosmological problems associated with $\phi$ coherent oscillation is avoided.

Note, however, that there is a subtlety in the model (\ref{toy_pot}).
After $S$ starts to oscillate, until its amplitude reduces to 
$A_S \sim m_\phi / \lambda$, the dynamics of $\phi$ is dominated by the interaction term $\lambda^2 S^2 \phi$ in the equation of motion (\ref{EOM_phi}).
Since the universe is assumed to be matter dominated, (\ref{EOM_phi}) is rewritten as  
\be
	\ddot{u} + \frac{\lambda^2 A_S^2}{2} \big[ 1+\cos (2m_S t) \big] u = 0,
\ee
where $u=(t/t_i) \phi$ with the initial time $t_i$ chosen freely.
This equation is mathematically equivalent to the Mathieu equation and there exists an exponentially increasing solution.\footnote{
We consider the resonant amplification of the homogeneous mode.
The finite wave number mode will also be amplified, but it does not much affect the following estimation of the $\phi$ energy density.
}
The behavior of the solution obeys the broad resonance regime discussed in Ref.~\cite{Kofman:1994rk}.
The number density of $u$ increases exponentially while the adiabatic condition is violated.
It takes place when $S$ crosses the origin,\footnote{
This is the case for a real scalar $S$. If $S$ is a complex scalar and $S^2$ in the potential (\ref{toy_pot}) is replaced with $|S|^2$, along with a certain magnitude of U(1) violating term $\propto (S^2 + S^{*2})$, the $S$ may rotate in the complex plane and not pass through the origin \cite{Allahverdi:1999je,Chacko:2002wr}. 
In this case, there are no such parametric resonant processes.}
or  $m_S t \sim (n + 1/2) \pi$ for an integer $n$, and the number density increases by a factor $\exp(2\pi \mu)$, where $\mu$ is a randomly-changing instability parameter and it takes $0.17$ on average.

\begin{figure}[t]
\centering
\subfigure[$S_i=400$]{
	\includegraphics [width = 7.5cm, clip]{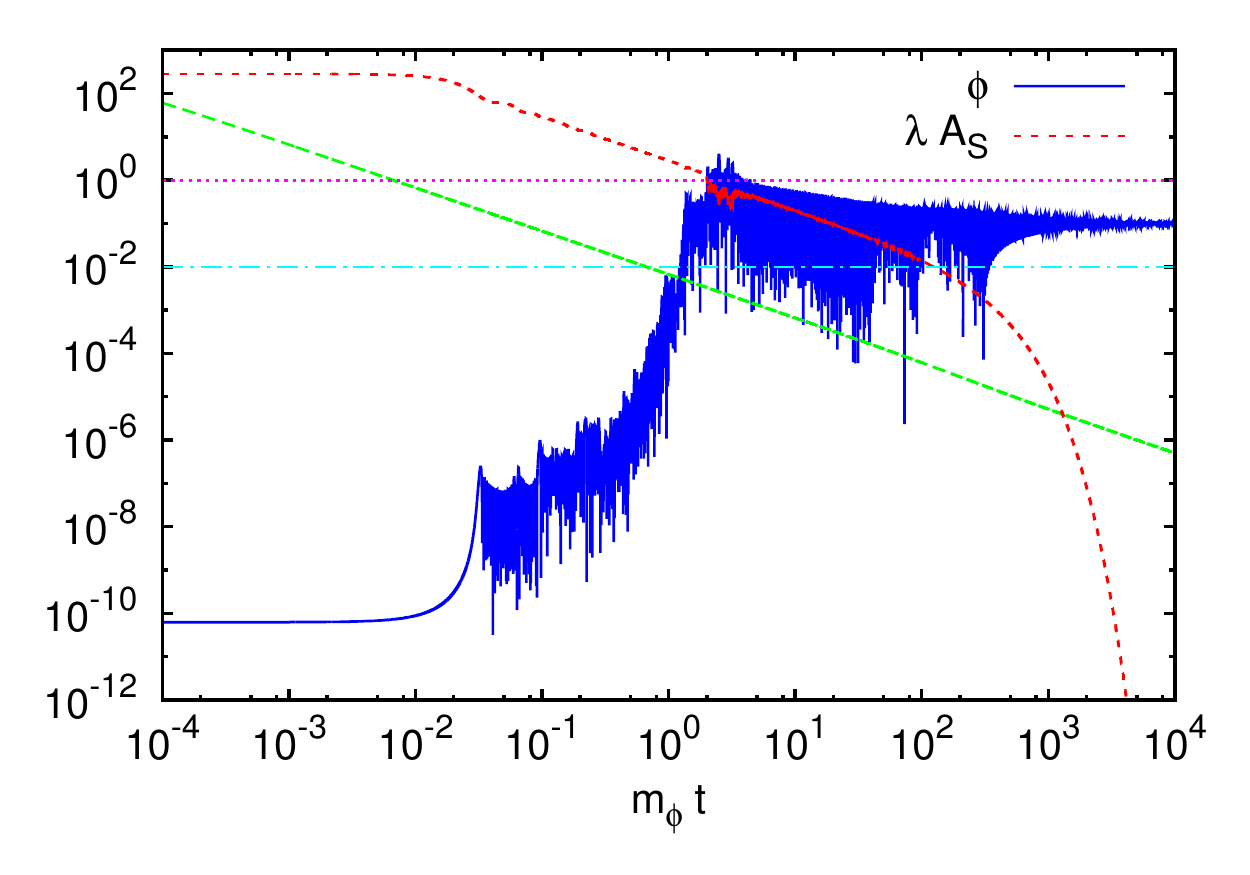}
	\label{Fig1a}
}
\subfigure[$S_i=50$]{
	\includegraphics [width = 7.5cm, clip]{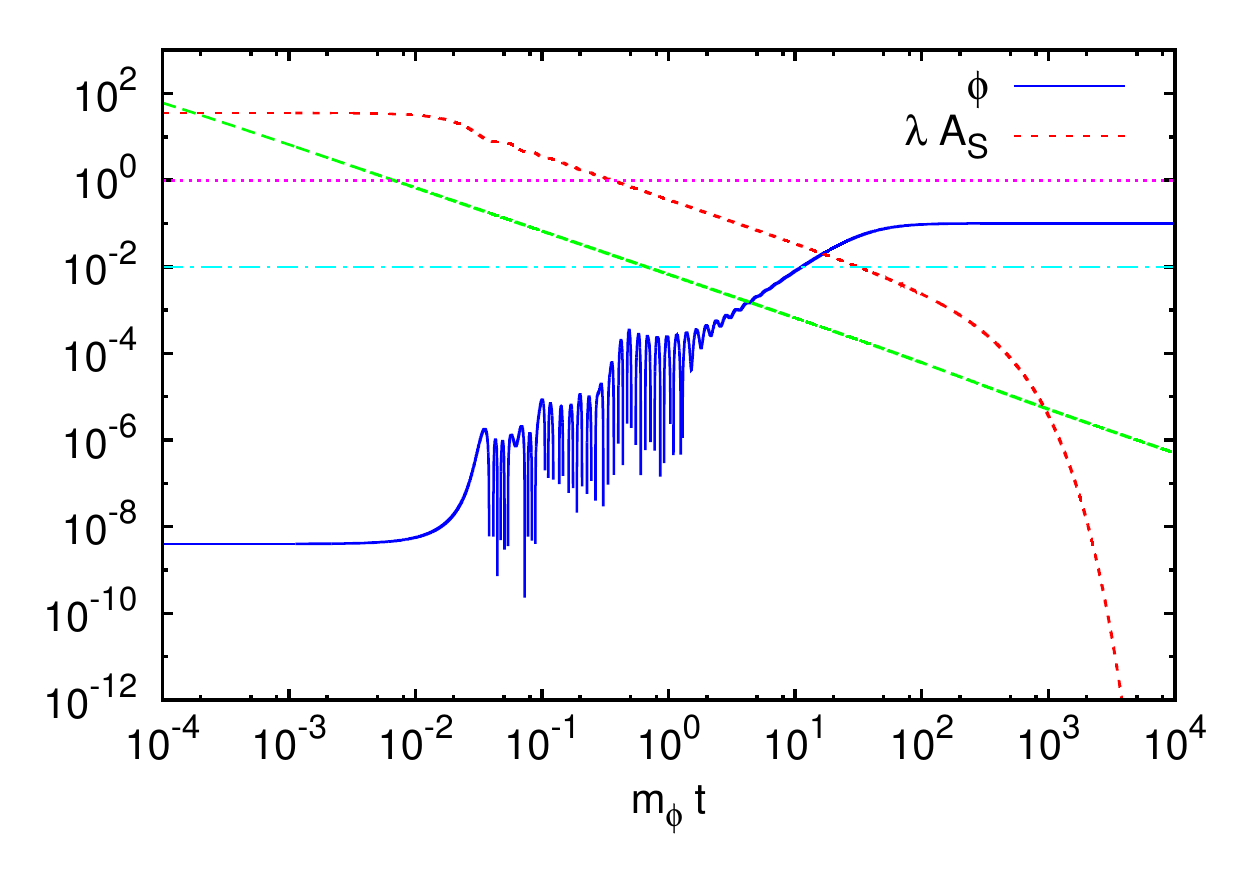}
	\label{Fig1b}
}
\caption{
Time evolutions of $\lambda A_S$ (red dotted line), $\phi$ (blue solid line), and $H$ (green dashed line) are shown.
The small dotted magenta line and the dashed-and-dotted cyan line correspond to $m_S$ and $m_\phi$ respectively.
Mass-dimensional values are normalized by some arbitrary mass scale.
We have taken $m_S=1,~ m_\phi = 0.01, ~\phi_0=0.1$, $\lambda = 1$, and $\Gamma_S = 10^{-4}$ (the decay rate of $S$) in both Figures and $S_i = 400$ in Fig.~\ref{Fig1a} and $S_i=50$ in Fig.~\ref{Fig1b}.
The initial amplitude of $\phi$ is set to be the minimum of the potential (\ref{phimin}).
}
\label{Fig1}
\end{figure}

The stochastic resonance continues as long as the frequency of $u$ is larger than that of $S$, and ceases at $A_S \sim m_S/\lambda \equiv S_{\rm end}$.
Note that, as is already mentioned, we need $S_i > S_{\rm end}$ in order for the adiabatic suppression to work.
If the number of $S$ oscillations during $S_i > A_S > S_{\rm end}$ is huge enough, the resonance continues until the produced $\phi$ energy density becomes comparable to that of $S$.
This happens at $\phi \sim m_S/\lambda\sim S_{\rm end}$, hence the $\phi$ amplitude may be raised up to that of $S$ in the limit of efficient parametric resonance.
This is shown in Fig.~\ref{Fig1a}, where we have solved the equations of motion (\ref{EOM_phi}) and (\ref{EOM_S}) and shown the time evolutions of $\phi$ and $S$ with following parameters : $m_S=1,~ m_\phi = 0.01, ~\phi_0=0.1$, $\lambda = 1$, $\Gamma_S = 10^{-4}$ and$S_i=400$ in arbitrary units.

On the other hand, the resonance ceases soon after $S$ starts to oscillate if the duration of the resonant oscillation is short.
This is the case if values of $S_i$ and $S_{\rm end}$ are rather close to each other.
In this case, the resulting amplitude of $\phi$ is highly suppressed compared with $A_S$. 
Fig.~\ref{Fig1b} shows the time evolutions of $\phi$ and $S$ with $m_S=1,~ m_\phi = 0.01, ~\phi_0=0.1$, $\lambda = 1$, $\Gamma_S = 10^{-4}$ and $S_i=50$ in arbitrary units.
It is clearly seen that the $\phi$ adiabatically follows the minimum of the potential without significant oscillations around the minimum although some amount of oscillations are induced via parametric resonance.

The amplitude of $\phi$ is denoted by $\phi_* =\xi S_{\rm end} = \xi m_S / \lambda$ at the time $A_S = S_{\rm end}=m_S/\lambda$, where $\xi$ represents a suppression factor which is determined by the ratio $S_i / S_{\rm end}$.
We have found that the suppression is most efficient ($\xi \ll 1$) for $S_i / S_{\rm end} \sim 10$ and $\xi=1$ for $S_i / S_{\rm end} \gtrsim 100$.

Let us briefly summarize necessary conditions for the adiabatic suppression to work.
\begin{itemize}
\item $S_i > S_{\rm end}\equiv m_S/\lambda$ : Otherwise, the effective mass for $\phi$ is smaller than the Hubble parameter and the adiabatic suppression does not work.
\item $S_i \ll 100S_{\rm end}$ : Otherwise, the resonant amplification of $\phi$ is efficient.
Combined with the above condition, the suppression is most efficient for $S_i \sim 10S_{\rm end}$.
\item $\Gamma_S < H_{\rm adi}\equiv m_\phi S_{\rm end}/S_i$ : Otherwise, $S$ decays and large effective mass for $\phi$ disappears before $\phi$ relaxes to the true minimum.
\end{itemize}

\subsection{The abundance of the oscillating scalar field}

We have seen that the effect of parametric resonance induces the amplitude of $\phi=\phi_*$ at $t=t_*$ where $t_* (\simeq \lambda S_i/m_S^2)$ is the cosmic time when $S$ becomes equal to $S_{\rm end}$.
Now let us see how the amplitude of $\phi$ decreases after that.
First we consider the dynamics of the fields in the era of $S_{\rm end} \gg A_S \gg m_\phi/\lambda$ (or $ m_S \gg \lambda A_S \gg m_\phi $).
Since the frequency of $S$ ($\sim m_S$) is much larger than that of $\phi$ ($\sim\lambda A_S$), 
we can replace the $S^2$ with its one cycle average $\langle S^2 \rangle = A_S^2/2$ which is proportional to $t^{-2}$.
Therefore the equation of motion of $\phi$ becomes a simple homogeneous equation and its solution is given by
\be
	\phi = \phi_* \bigg( \frac{t}{t_*} \bigg)^{-1/2} 
	\cos \bigg[ 
	\frac{\sqrt{2} \lambda A_S }{3 H} \log \bigg( \frac{t}{t_*} \bigg) 
	\bigg].
\ee
Note that the amplitude of the $\phi$, $A_\phi$, is damped as $A_\phi \propto t^{-1/2}$ in this era.
Next, when $\lambda A_S \ll m_\phi$, the bare mass for $\phi$ dominates and the solution of the equation of motion is expressed as
\be
	\phi = \phi_0 + \frac{B}{t} \cos (m_\phi t),
\ee
where $B$ is a constant of integration.
This shows that the amplitude of the oscillation of $\phi$ around the true minimum decreases as $A_\phi \propto t^{-1}$.

\begin{figure}[t]
\centering
\subfigure[$S_i = 100$]{
	\includegraphics [width = 7.5cm, clip]{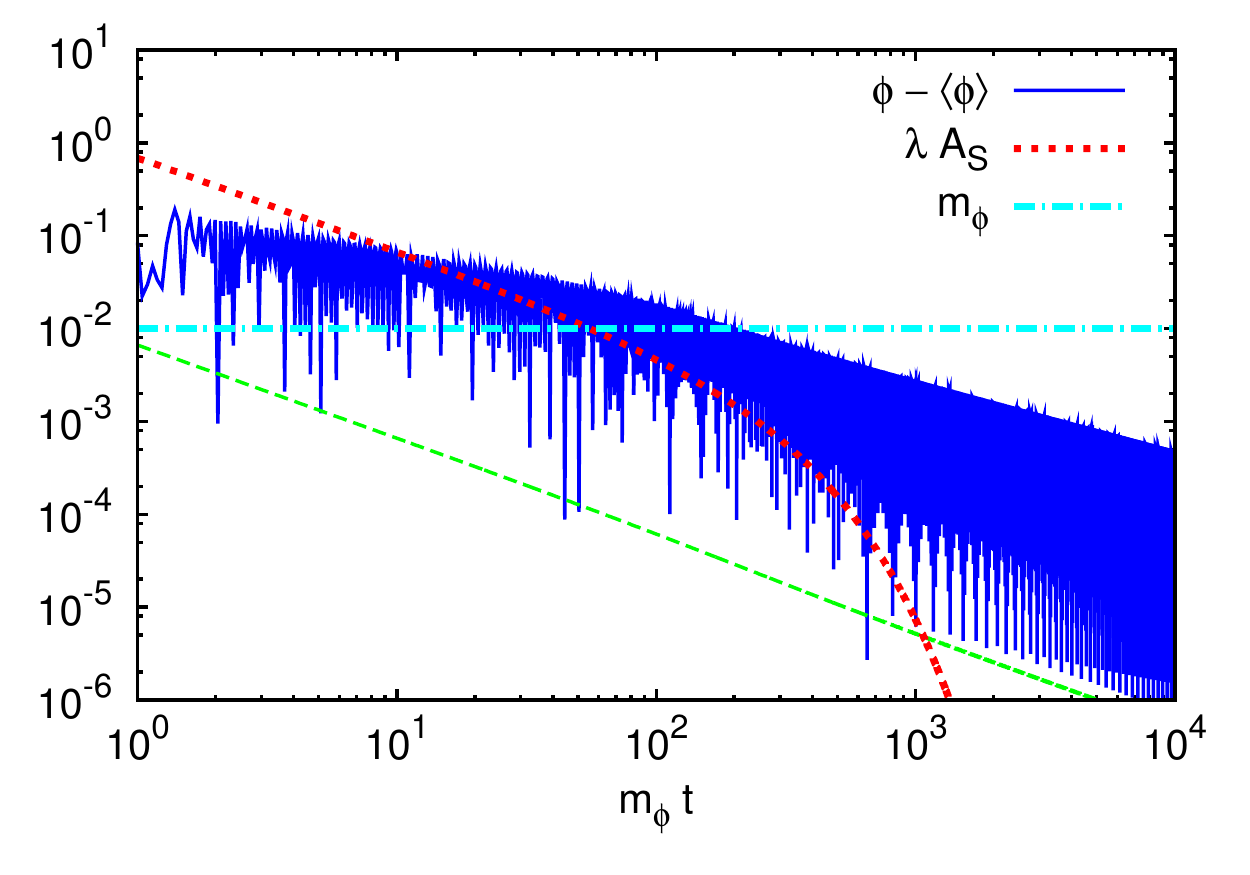}
	\label{Fig2a}
}
\subfigure[$S_i=0.1$]{
	\includegraphics [width = 7.5cm, clip]{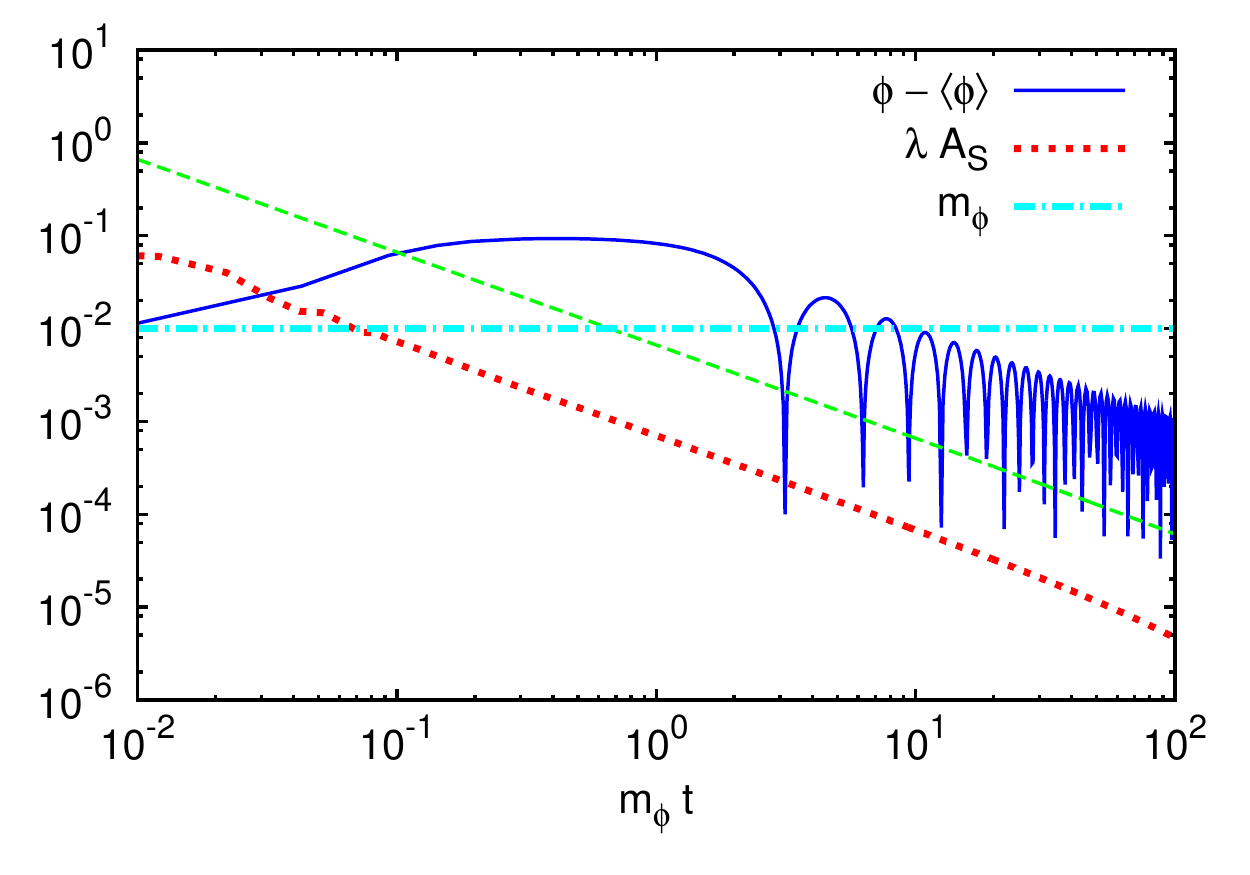}
	\label{Fig2b}
}
\caption{
Time evolutions of $\lambda A_S$ (dotted red line), $H$ (dashed green line), and $\phi - \langle \phi \rangle$ (solid blue line) are shown.
The dashed-and-dotted cyan line represents $m_\phi$.
Parameters are the same as those in Fig.~\ref{Fig1} but the initial amplitude of $S$ are $S_i = 100$ in Fig.~\ref{Fig2a} and $S_i=0.1$ in Fig.~\ref{Fig2b}.
}
\label{Fig2}
\end{figure}

These results are supported by the numerical calculation shown in Fig.~\ref{Fig2a}, where the time evolutions of $\phi$ and $S$ are shown for $S_i=100$ with other parameters taken to be the same as those in Fig.~\ref{Fig1}.
From the figure, we can see that the center of the $\phi$ oscillation traces the time-dependent minimum of the potential adiabatically.
This behavior cannot be realized if $\lambda S_i < m_S$, in which the minimum changes while $\phi$ is frozen and $\phi$ starts to oscillate at $H = m_\phi$ with initial amplitude $\phi_0$, as shown in Fig.~\ref{Fig2b} where we have taken $S_i=0.1$.

Now we can calculate the abundance of the $\phi$ oscillation.
As we have explained we assume that $S$ as well as the inflaton decays much later.
In this case, the dynamics is qualitably same as that in Fig.~\ref{Fig2a}.
In order to estimate the $\phi$ abundance, notice that the comoving $\phi$ number density after the resonance ends is adiabatically invariant. 
Thus we have $n_\phi(t) = n_\phi(t_*)[a(t_*)/a(t)]^3$ where $n_\phi(t_*)=\lambda S_{\rm end} \phi_*^2/2$.
By noting that the final energy density of $\phi$ around the true minimum is given by $\rho_\phi(t)=m_\phi n_\phi(t)$, we obtain the following $\rho_\phi$ to entropy ratio,
\be
	\frac{\rho_\phi}{s} 
	= \frac{1}{8} T_R \bigg(\frac{\xi^2 m_\phi}{m_S} \bigg) 
	\bigg( \frac{S_i}{M_P} \bigg)^2 
	= \frac{1}{8} T_R \gamma \bigg( \frac{\phi_0}{M_P} \bigg)^2,
\ee
where $\gamma$ denotes the suppression factor compared with the case in which the adiabatic suppression does not occur, and it is given by 
\be
	\gamma \equiv \frac{\xi^2m_\phi}{m_S} \bigg( \frac{S_i}{\phi_0} \bigg)^2.
	\label{gamma_toy_model}
\ee
Hence the adiabatic suppression is efficient if $\gamma \ll 1$.

\begin{figure}[t]
\centering
\subfigure[]{
	\includegraphics [width = 7.5cm, clip]{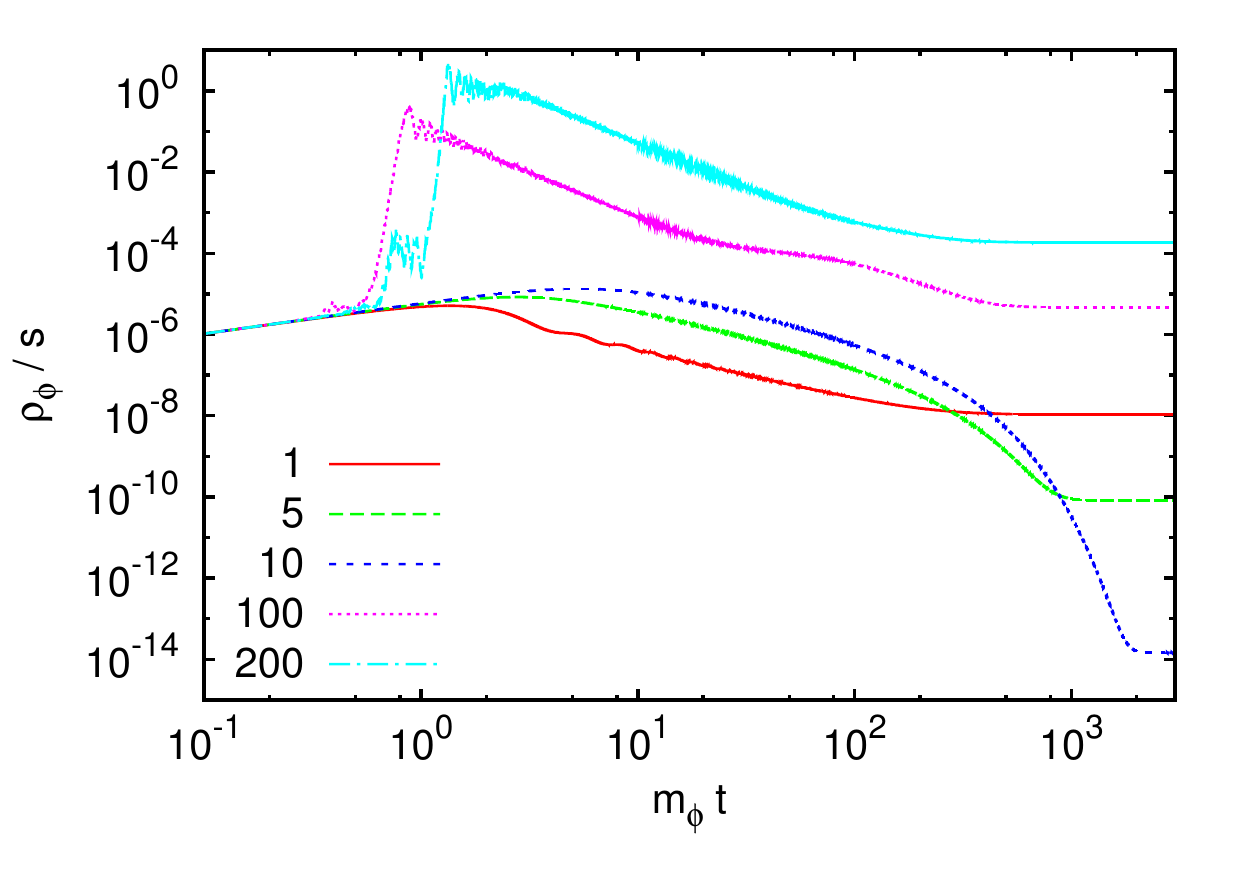}
	\label{Fig3a}
}
\subfigure[]{
	\includegraphics [width = 7.5cm, clip]{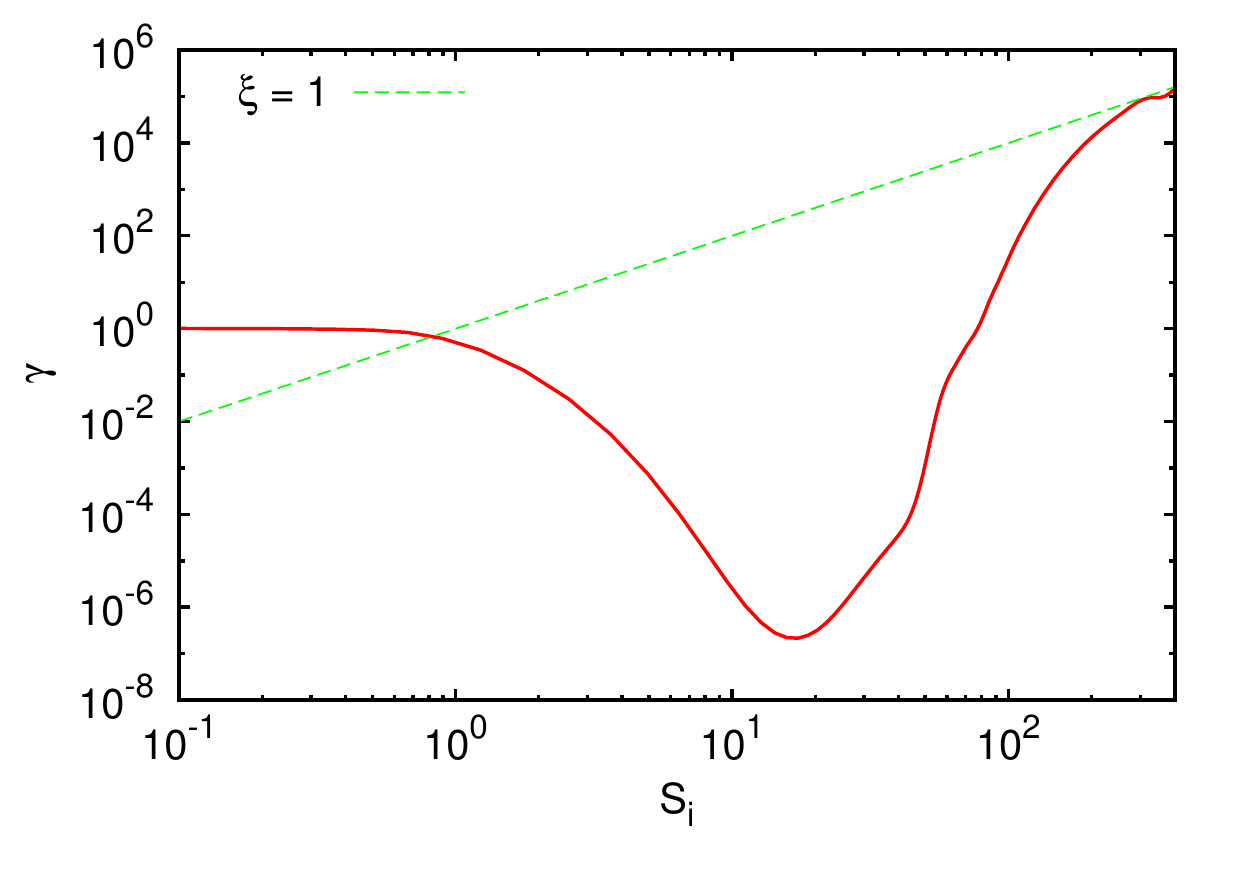}
	\label{Fig3b}
}
\caption{
Time evolutions of $\rho_\phi / s$ (left) and $S_i$ dependence of the suppression factor $\gamma$ (right) are shown.
Mass dimensional values are normalized by $M=0.01M_P$.
We have taken $m_S=1$, $m_\phi=0.01$, $\phi_0=0.1$, $\lambda = 1$, and $\Gamma_S = 10^{-4}$. 
In the left figure, we have taken $S_i = 1$ (solid red line), $S_i=5$ (dashed green line), $S_i=10$ (dotted blue line), $S_i=100$ (small-dotted magenta line), and $S_i=200$ (dashed-and-dotted cyan line).
In the right figure, the dashed green line represents the analytical formula given by (\ref{gamma_toy_model}) with $\xi = 1$.
}
\label{Fig3.1}
\end{figure}

We calculated numerically the time evolution of the energy-to-entropy ratio until it is fixed at the reheating and the result is shown in Fig.~\ref{Fig3a}.
We also show the $S_i$ dependence of the suppression factor $\gamma$ 
in Fig.~\ref{Fig3b}.
We have taken $m_S=1$, $m_\phi=0.01$, $\phi_0=0.1$, $\lambda = 1$, and $\Gamma_S=10^{-4}$.
We found that if $S_i \gtrsim 200$, the resonance becomes efficient and $\xi \sim 1$.
The resonance is not efficient (i.e. $\xi \ll 1$) for $S_i \sim 10$, where the suppression mechanism works most efficiently.
In contrast, for $S_i \lesssim 1$, the suppression mechanism no longer works.
Note that the efficiency of the suppression depends on $S_i / S_{\rm end}$.
Therefore, for different choice of parameters, different values of $S_i$ are favored from the viewpoint of the adiabatic suppression mechanism.

\section{Application to a supersymmetric axion model} \label{axion}

The supersymmetric (SUSY) axion model is introduced to solve the problems of the standard model (SM) of particle physics.
One of the serious problem is known as the strong CP problem.
Quantum chromodynamics (QCD) allows the existence of the CP violating term in the Lagrangian, but on the other hand, the measurement of the neutron electric dipole moment shows that CP must be preserved with high accuracy~\cite{Baker:2006ts}.
This implies that the CP violating term must be highly suppressed, and the SM cannot explain this fact in a natural way.
The most popular solution was proposed by Peccei and Quinn~\cite{Peccei:1977hh}.
They introduced an additional global $U(1)$ symmetry, called Peccei-Quinn (PQ) symmetry, written as $U(1)_\mathrm{PQ}$.
When the PQ symmetry is broken spontaneously, the axion arises as a pseudo Nambu-Goldstone boson~\cite{Kim:1986ax}.
The axion acquires its mass through the QCD instanton effect and settles down to the CP preserving minimum of the potential.
Another problem of the SM is the gauge hierarchy problem.
It is naturally solved in the framework of supersymmetry (SUSY)~\cite{Martin:1997ns}.
Therefore, the SUSY axion model solves both of the problems in the SM.

In the SUSY axion model, the scalar partner of the axion, saxion, and the fermionic superpartner of the axion, axino, are included in the PQ supermultiplet and they take significant roles in cosmology~\cite{Rajagopal:1990yx,Kim:1992eu,Lyth:1993zw,Hashimoto:1998ua,Asaka:1998ns,Banks:2002sd,Kawasaki:2007mk,Kawasaki:2008jc}.
In this section, we focus on the dynamics of the saxion and verify that the mechanism shown in the previous section is applicable.
This was partly mentioned in Ref.~\cite{Kawasaki:2010gv} in the context of hybrid inflation model in the SUSY axion model.
We further investigate this issue in a more general form.

\subsection{The potential of the SUSY axion model}

Let us consider the following superpotential,
\be
	W = \kappa S (\Psi \bar\Psi - f_a^2),  \label{superpot}
\ee
where $S$ is a gauge singlet superfield and has a zero PQ charge and $\Psi$ and $\bar\Psi$ are the gauge singlet PQ superfields whose PQ charges are $+1$ and $-1$, respectively.
The PQ superfields contain the axion ($a$), saxion ($\sigma$) and axino ($\tilde a$).
Here $f_a$ is the PQ symmetry breaking scale and $\kappa$ is dimensionless coupling constant which is assumed to be real and positive.
The scalar potential derived from the superpotential (\ref{superpot}) is
\be
	V = \kappa^2 |\Psi \bar\Psi - f_a^2|^2 + \kappa^2 |S|^2 ( |\Psi|^2 + |\bar\Psi|^2 ),	\label{PQ_pot}
\ee
where we used the same symbol for the superfield and its scalar component.
The global minimum lies at $S=0$ and $\Psi\bar\Psi = f_a^2$.
The flat direction along $\Psi\bar\Psi = f_a^2$ is stabilized at $|\Psi|\sim |\bar\Psi|\sim f_a$ by including the SUSY breaking effect and correspondingly the saxion, the flat direction in the scalar potential of $\Psi$ and $\bar\Psi$, obtains a mass of order of the gravitino, $m_{3/2}$.
Note that for $|S| > f_a$, the minimum of the PQ scalar $\Psi~(\bar\Psi)$ is $\Psi=0~(\bar\Psi=0)$, so the PQ symmetry is restored.
For $|S| < f_a$ on the other hand, the PQ scalars settle on the flat direction $\Psi \bar\Psi = f_a^2$ and the PQ symmetry is spontaneously broken.
In order to apply the adiabatic suppression mechanism to the SUSY axion model, we assume that $S$ is initially displaced far from the origin.
In order to realize such a condition, for example, we add a negative Hubble-induced mass term of $S$ to the potential, which comes from the supergravity effect~\cite{Dine:1995kz}, and introduce an additional nonrenormalizable superpotential
\be
	W_\mathrm{NR} = \frac{XS^{n-1}}{M^{n-3}}, \label{NR_superpot}
\ee
where $X$ is an additional superfield and $M$ is a cutoff scale and $n$ is a positive integer $(n \geq 4)$.
It is achieved by assigning appropriate discrete $R$-charges on $S$ and $X$.
Then $S$ obtains the potential
\be
	V_S = -c_H H^2 |S|^2 + \frac{|S|^{2(n-1)}}{M^{2(n-3)}}, \label{S_pot}
\ee
where we assume that $X$ sits at the origin due to the positive large Hubble-induced mass.
Hence the $S$ tracks the temporal minimum given by $|S| \sim (HM^{n-3})^{1/(n-2)}$, which is displaced far from the origin.
According to the potential (\ref{PQ_pot}), the PQ fields $\Psi$ and $\bar\Psi$ have the same mass $m_\Psi \sim \kappa |S|$ soon after the PQ symmetry breaking.
Hence the expectation values during this period are given by $\langle \Psi \rangle = \langle \bar\Psi \rangle = f_a$.
The condition that the effective mass of the PQ field must be larger than the Hubble parameter is $\kappa f_a > H_{\rm PQ}$ where $H_{\rm PQ}$ denotes the Hubble parameter at the beginning of $S$ oscillation induced by the PQ symmetry breaking.\footnote{
	For $n=4$, it is estimated as $H_{\rm PQ} \sim f_a^2/M$ for $f_a \ll \kappa M$
	and  $H_{\rm PQ} \sim \kappa f_a$ for $f_a \gg \kappa M$.
}
On the other hand, the condition that the energy density of the Universe is dominated by the inflaton at the PQ symmetry breaking is written as $\kappa^2 f_a^4 < 3M_P^2 H_{\rm PQ}^2$.
These conditions are rewritten as $f_a/M < \kappa  < \sqrt{3}M_P/M$ by considering the superpotential (\ref{NR_superpot}) with $n=4$.
Therefore, for a natural value of $\kappa$, the adiabatic suppression is realized.\footnote{
If $S$ dominates the Universe, inflation takes place and ends at $S=f_a$ at which tachyonic instability along the direction of PQ scalars develops.
This is the hybrid inflation model described in Ref.~\cite{Kawasaki:2010gv}.}

The dynamics is somewhat complicated. 
Let us briefly summarize the dynamics before going into details described in the following subsections.
\begin{itemize}
\item First, $S$ has a large value due to the potential (\ref{S_pot}). 
PQ scalars are stabilized at the origin $\Psi=\bar\Psi=0$ due to the large mass induced by $S$. $S$ gradually decreases due to the Hubble expansion.
\item At $H=H_{\rm PQ}$, where $S \sim f_a$, the tachyonic instability develops for PQ scalars and they fall down to the flat direction $\Psi\bar\Psi = f_a^2$. 
\item Almost simultaneously, the $S$ field begins to feel large mass from the coupling to $\langle\Psi\rangle$ and $\langle\bar\Psi\rangle$, and then $S$ begins to oscillate around the minimum $S=0$ with initial amplitude of $S\sim f_a$.
\item Then, the system resembles the toy model in the previous section.
The saxion, corresponding to the flat direction in the $\Psi$ and $\bar\Psi$ space, obtains a large mass from the oscillating scalar $S$, and it follows the temporal minimum of the potential if the following condition is satisfied : $\kappa f_a \gg H_{\rm PQ}$.
\item The parametric resonance occurs soon after the $S$ starts oscillates and it ends within one Hubble time.
During this short period, however, the resonance effect may efficiently amplify the saxion abundance since the oscillation frequency is much larger than the Hubble scale.
\item After the resonance ends, the saxion adiabatically follows the minimum until it relaxes to the true minimum when $S$ reduces to $S \sim S_{\rm adi}\equiv m_{3/2}/\kappa$ or $H\sim H_{\rm adi}\equiv H_{\rm PQ}m_{3/2}/(\kappa f_a)$, if $S$ decays much later.
\end{itemize}


\subsection{The dynamics of the saxion}

\subsubsection{Soon after the PQ symmetry breaking}

\begin{figure}[t]
\centering
\subfigure[$\kappa = 1$]{
	\includegraphics [width = 7.5cm, clip]{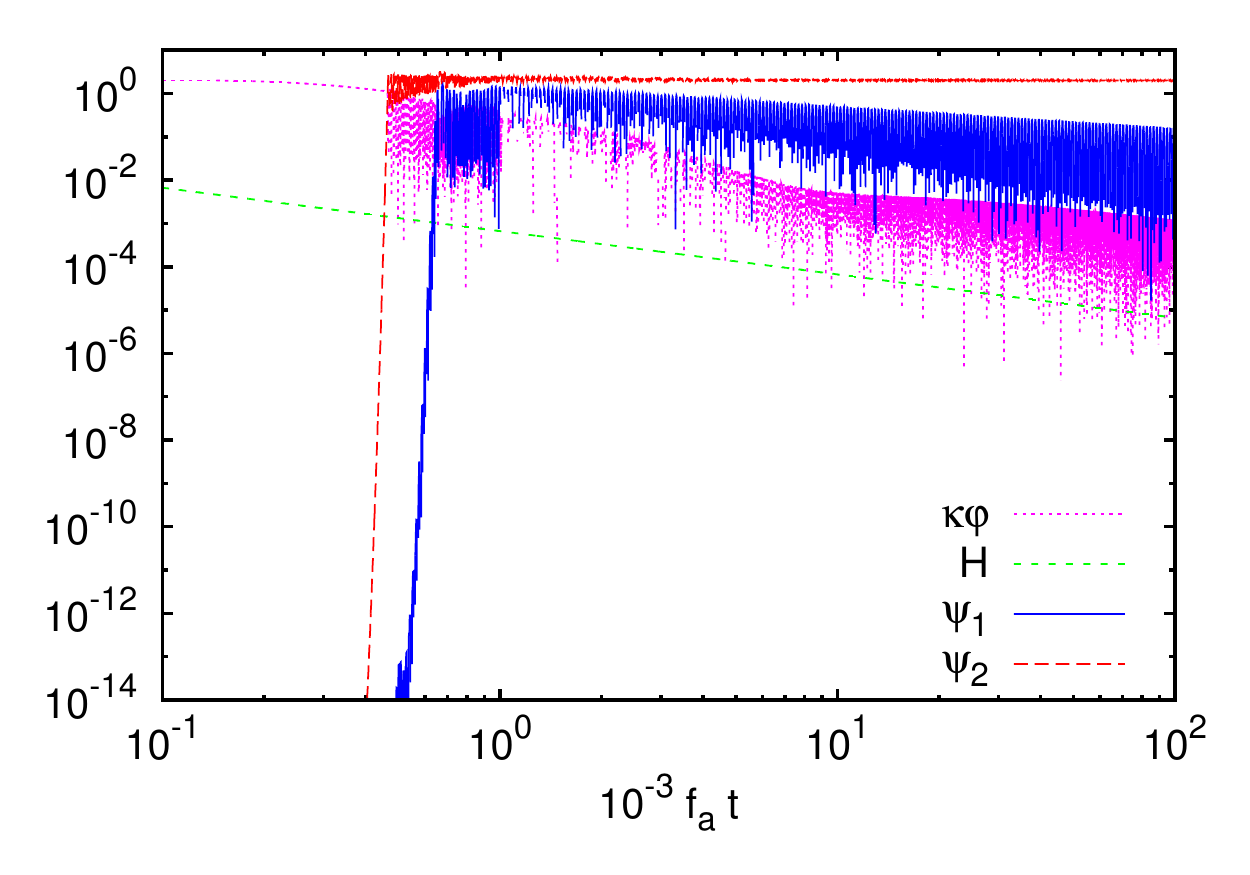}
	\label{Fig4a}
}
\subfigure[$\kappa=0.01$]{
	\includegraphics [width = 7.5cm, clip]{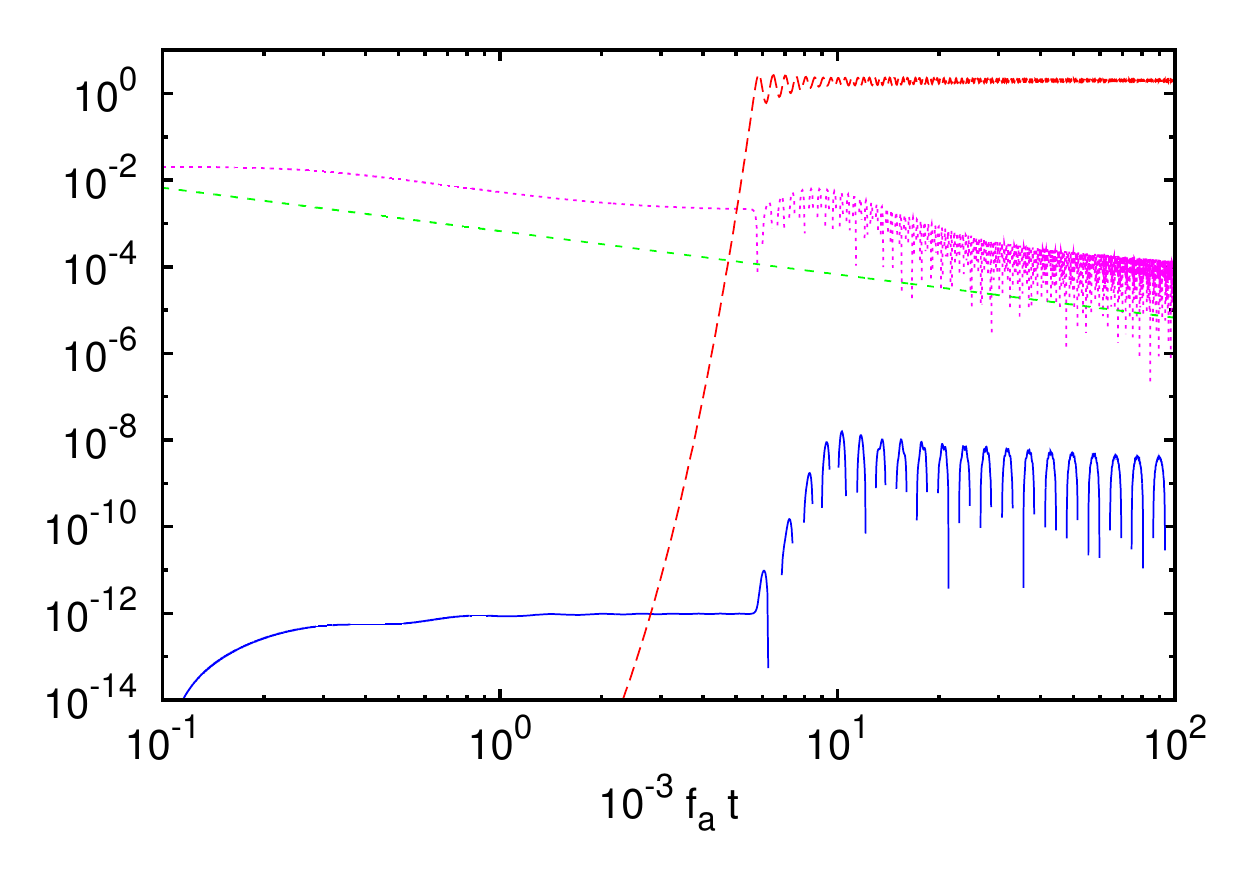}
	\label{Fig4b}
}
\subfigure[$\kappa=0.001$]{
	\includegraphics [width = 7.5cm, clip]{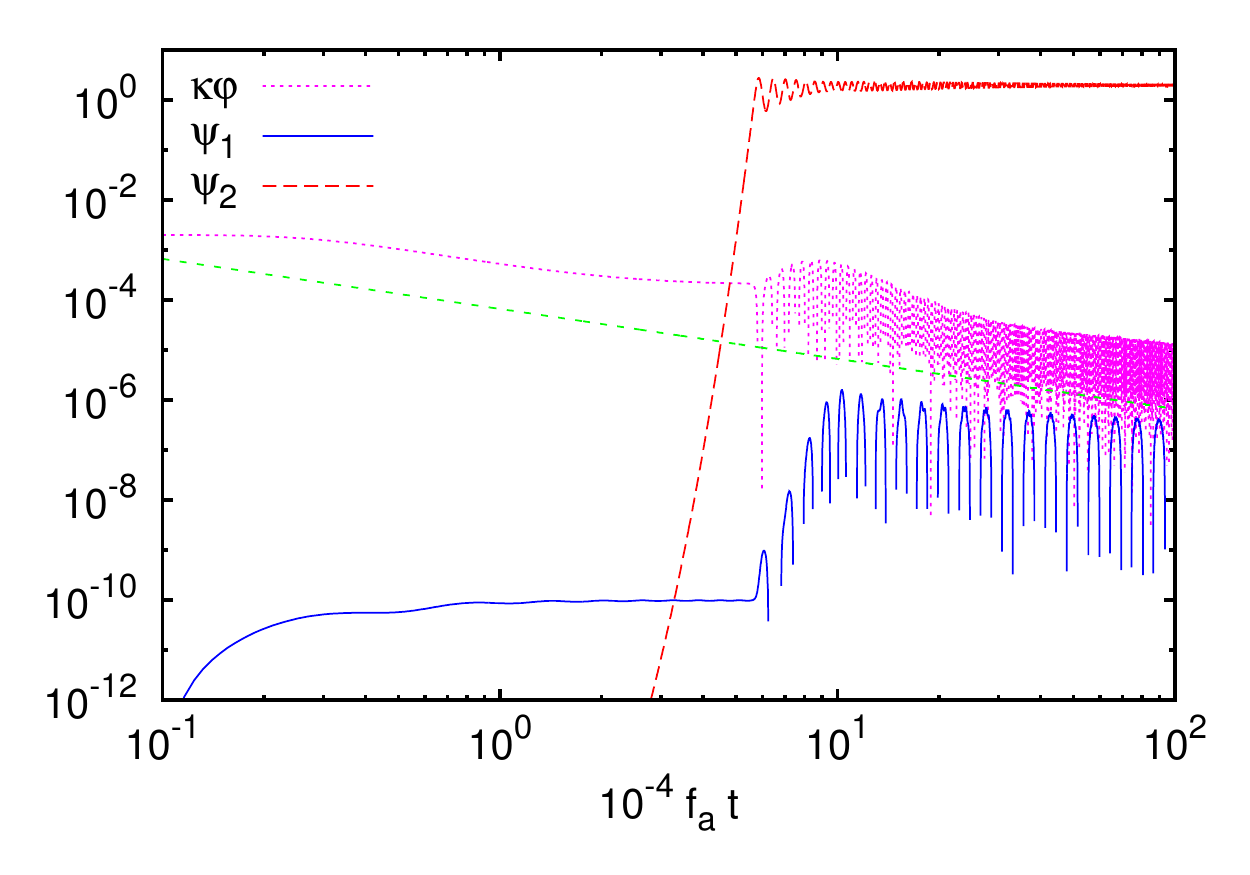}
	\label{Fig4c}
}
\caption{
Time evolutions of $\kappa \varphi$ (small dotted magenta line), $\psi_2$ (dashed red line), and $\psi_1$ (solid blue line) are shown.
Mass dimension is normalized by the PQ scale $f_a$.
We have taken $\varphi = 2$ and $\psi_1 = \psi_2 = 10^{-20}$ as initial values 
and $m_{3/2}=10^{-8}$, $n=4$, $c_H=1$ and
(a) : $H=6\times 10^{-3}$, $M=5 \times 10^2$, $\kappa = 1$,
(b) : $H=6\times 10^{-3}$, $M=5 \times 10^2$, $\kappa = 0.01$,
(c) : $H=6\times 10^{-4}$, $M=5 \times 10^3$, $\kappa = 0.001$, where $H$ is the initial Hubble parameter.
}
\label{Fig4}
\end{figure}

First, we consider the PQ field dynamics in some short epoch after the PQ symmetry breaking, in which the low energy SUSY breaking terms are assumed to be completely negligible : $\kappa S \gg m_{3/2}$.
We denote the flat direction and the direction perpendicular to it as $\Psi_1$ and $\Psi_2$, respectively.These directions are given by
\be
	\Psi_1 = \frac{\Psi - \bar\Psi^*}{\sqrt{2}},
	~ \Psi_2 = \frac{\Psi^* + \bar\Psi}{\sqrt{2}},
\ee
and the potential of the PQ fields (\ref{PQ_pot}) is rewritten as
\be
	\begin{split}
		V &= \kappa^2 f_a^4 + \kappa^2 (|S|^2 + f_a^2)|\Psi_1|^2 
		+ \kappa^2 (|S|^2 - f_a^2) |\Psi_2|^2 \\[1mm]
		&~~~ +\frac{\kappa^2}{4}|\Psi_1|^4 + \frac{\kappa^2}{4}|\Psi_2|^4
		-\frac{\kappa^2}{2}|\Psi_1|^2 |\Psi_2|^2 [1-2\sin^2(\theta_1+\theta_2)] 
		+ V_S,
	\end{split}
\ee
where $\theta_1=\mathrm{arg}(\Psi_1)$ and $\theta_2=\mathrm{arg}(\Psi_2)$.
Assuming that $\Psi_1 = \Psi_1^*$, $\Psi_2 = \Psi_2^*$, and $S = S^*$ and defining $\psi_1 = \sqrt{2} \Psi_1$, $\psi_2 = \sqrt{2} \Psi_2$, and $\varphi = \sqrt{2} S$,  the potential is rewritten as
\be
	V=\kappa^2 f_a^4 + \frac{1}{2}\kappa^2 f_a^2 (\psi_1^2-\psi_2^2) 
	+ \frac{1}{4}\kappa^2 \varphi^2 (\psi_1^2 + \psi_2^2) 
	+ \frac{1}{16} \kappa^2 \psi_1^4 + \frac{1}{16} \kappa^2 \psi_2^4
	-\frac{1}{8} \kappa^2 \psi_1^2 \psi_2^2  + V_S.
\ee
From this potential, the equations of motion of respective directions are given by 
\be
	\ddot{\varphi} + 3H \dot{\varphi} 
	+ \frac{\kappa^2}{2} (\psi_1^2 + \psi_2 ^2) \varphi -c_H H^2 \varphi 
	+ \frac{3 \varphi^5}{4M^2} = 0,   \label{EOM_varphi}
\ee
\be
	\ddot{\psi_1} + 3H \dot{\psi_1} 
	+ \kappa^2 \bigg( f_a^2 + \frac{\varphi^2}{2} - \frac{\psi_2^2}{4} \bigg) \psi_1 
	+ \frac{\kappa^2}{4} \psi_1^3 = 0,   \label{EOM_phi1}
\ee
\be
	\ddot{\psi_2} + 3H \dot{\psi_2} 
	+ \kappa^2 \bigg( - f_a^2 + \frac{\varphi^2}{2} -\frac{\psi_1^2}{4} \bigg) \psi_2 
	+ \frac{\kappa^2}{4} \psi_2^3 = 0,   \label{EOM_phi2}
\ee
where we have taken $n=4$.
Since after the PQ symmetry breaking the PQ scalars have $\langle \Psi \rangle = \langle \bar\Psi \rangle \simeq f_a$, the expectation values of respective directions are $\langle \psi_1 \rangle \simeq 0$ and $\langle \psi_2 \rangle \simeq 2 f_a$.
By looking at the mass terms for $\psi_1$ and $\psi_2$, it is clear that the $\psi_2$ starts to oscillate with initial amplitude of $\sim f_a$ around its minimum after the PQ breaking, while the flat direction $\psi_1$ remains at the origin.
The approximate solutions of equations (\ref{EOM_varphi}) and (\ref{EOM_phi2}) are written as
\be
	\varphi \simeq \frac{f_a}{\sqrt 2} \bigg( \frac{t_i}{t} \bigg) \cos [ 2 \kappa f_a (t-t_i) ] , ~~ 
	\psi_2 \simeq 2 f_a +2  f_a \bigg( \frac{t_i}{t} \bigg) \cos [ 2 \kappa f_a (t-t_i)].
\ee
Looking at the equation of motion of the flat direction (\ref{EOM_phi1}), there exists a interaction term which may induce the parametric resonance, as shown in the previous section.
Since $\varphi \sim f_a$ soon after the PQ breaking, the frequency of the oscillation $\psi_1$ is almost equal to that of $\psi_2$, so the parametric resonance takes place within a short epoch after the PQ breaking.
However, the frequency itself is much larger than the Hubble scale at that epoch ($\kappa f_a \gg H_{\rm PQ}$), so there may be many oscillations during this short period.
Thus we cannot necessarily neglect the effect of parametric resonance.\footnote{
	As often discussed in the hybrid inflation scenario, the tachyonic preheating may be significant at the PQ symmetry breaking \cite{Felder:2000hj}.
	However, now we only focus on the flat direction, $\psi_1$ which does not become tachyonic at the PQ breaking.
	Since we will not  follow in detail the dynamics of the massive field, $\psi_2$ which becomes tachyonic, we neglect the effect of the tachyonic preheating in our context.}
To verify this, we have numerically solved the set of equations of motion (\ref{EOM_varphi})-(\ref{EOM_phi2}).
The numerical results are shown in Fig.~\ref{Fig4} and Fig.~\ref{Fig4c}, where time evolutions of $\kappa \varphi$ (small dotted magenta line), $\psi_2$ (dashed red line), and $\psi_1$ (solid blue line) are shown.
Behavior of $\varphi$ and $\psi_2$ shows the good agreement with the results in \cite{GarciaBellido:1997wm} 
in which the scalar field dynamics under the similar setup is investigated in detail.
We have taken $\varphi = 2$, $H=6\times 10^{-3}$, and $\psi_1 = \psi_2 = 10^{-20}$ as initial values 
and $\kappa = 1$ for Fig.~\ref{Fig4a} and $\kappa = 0.01$ for Fig.~\ref{Fig4b}.
In fact, the initial value of $\psi_1$ is not reflected unless it is too large because the initial amplitude of $\psi_1$ is determined by the tiny deviation from $\Psi = \bar\Psi = f_a$ at the PQ breaking, which is induced by the gravitino mass terms.
The resonance is efficient only if the Hubble parameter is much less than $\kappa f_a$ at the beginning of the oscillation of $\varphi$ and $\psi_2$.
We investigated it numerically and concluded that if $H_{\rm PQ} \lesssim 10^{-2} \kappa f_a$, the amplitude of the flat direction is raised up to $f_a$.
In the case of $H_{\rm PQ} \sim 10^{-1} \kappa f_a$, the resonant amplification of $\psi_1$ becomes inefficient much sooner due to the expansion of the Universe, and the amplitude of the flat direction is highly suppressed as shown in Fig.~\ref{Fig4b}.
The saxion is identified as a oscillation along the flat direction, so the initial amplitude of the saxion can be expressed as $\sigma_i = \xi f_a$ with $\xi \leq 1$.
In particular, we found $\xi \sim 10^{-8}$ in the case of Fig.~\ref{Fig4b} ($ \kappa f_a / H_{\rm PQ} \sim 10$ at the beginning of $S$ oscillation and $\kappa = 0.01$), 
and $\xi \sim 10^{-6}$ in the case of Fig.~\ref{Fig4c} ($ \kappa f_a / H_{\rm PQ} \sim 10$ at the beginning of $S$ oscillation and $\kappa = 0.001$).

Before ending this subsection, we comment on the parameters chosen in our numerical calculations.
As we will discuss later, a relatively large coupling constant $\kappa$ 
leads to too early decay of $S$ into the axino, which spoils the adiabatic suppression (see Eq.~(\ref{cond-m_3/2})).
In Fig.~\ref{Fig4c}, we have chosen parameters that satisfy Eq.~(\ref{cond-m_3/2}).
Although it is not satisfied for parameters in Fig.~\ref{Fig4b},
we have demonstrated the scalar dynamics just to show the validity of the arguments in the previous subsection.
It is clear that the adiabatic suppression occurs if the parameter choices are extrapolated to those satisfying Eq.~(\ref{cond-m_3/2}).

\subsubsection{Well after the PQ symmetry breaking}

\begin{figure}[t]
\centering
\includegraphics [width = 10cm, clip]{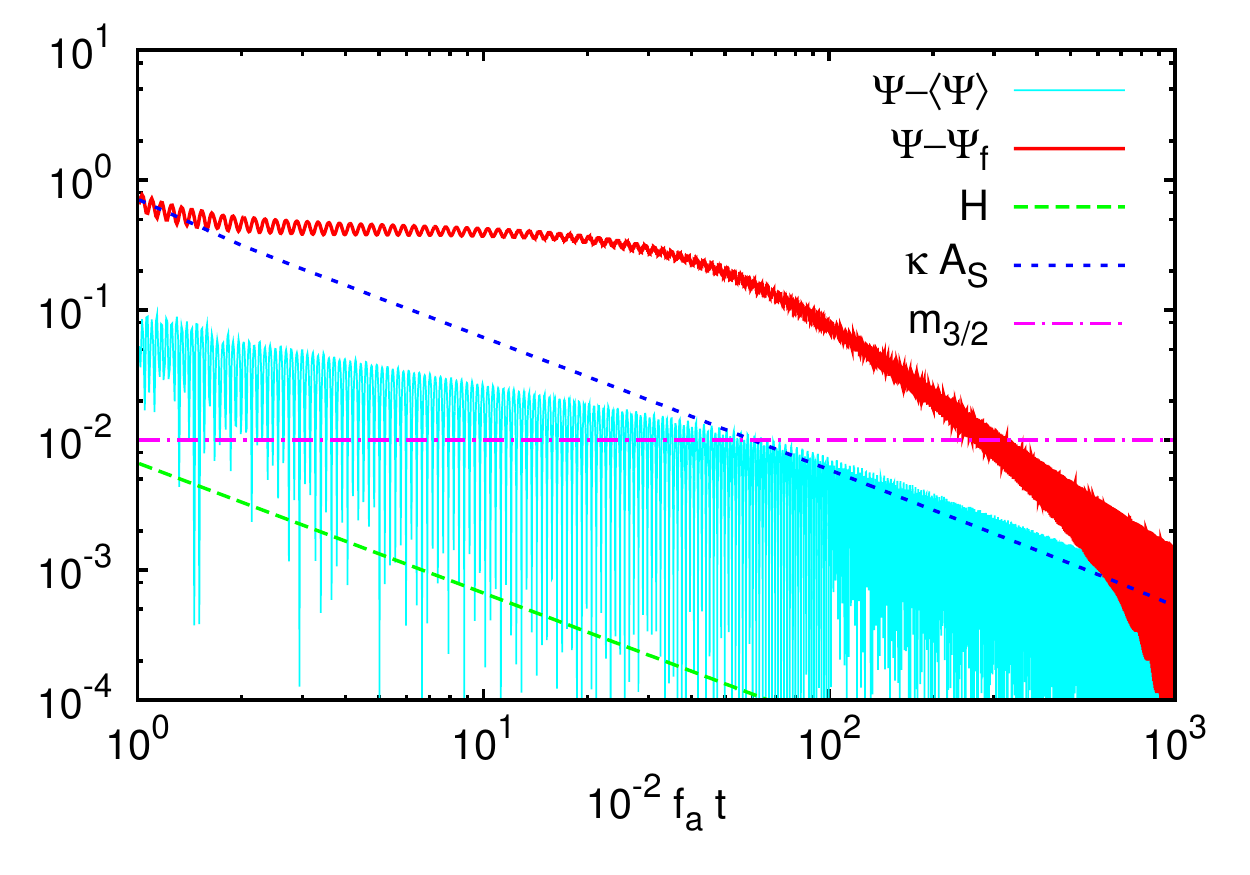}
\caption{
The time evolutions of $\Psi - \langle \Psi \rangle$ (displacement from the temporal minimum, thin solid cyan line), $\Psi - \Psi_f$ (displacement from the eventual minimum, thick solid red line), and $\kappa A_S$ (dotted blue line).
The Hubble parameter and the gravitino mass are represented as dashed green line and dashed-and-dotted magenta line respectively.
Mass scale is normalized by $f_a$.
We have taken $S=1$, $\Psi=0.1+\langle \Psi \rangle$, and $H=0.006$ as initial values and $\kappa = 1$ and  $m_{3/2}=0.01$
}
\label{Fig5}
\end{figure}

As the Universe is cooled, the low scale SUSY breaking terms become dominant. 
It happens at $S \lesssim S_{\rm adi}\equiv m_{3/2}/\kappa$ when the Hubble parameter becomes 
smaller than $H_{\rm adi}\equiv H_{\rm PQ}m_{3/2}/(\kappa f_a)$.
Then, the dominant contribution to the potential of the PQ fields along the flat direction is 
\be
	V=\kappa^2|S|^2(|\Psi|^2 + |\bar\Psi|^2) 
	+ m_{3/2}^2(c_1 |\Psi|^2 + c_2|\bar\Psi|^2) 
	~~\text{with}~~ \Psi\bar\Psi = f_a^2,
\ee
where $c_1$ and $c_2$ are assumed to be real and positive constants of order unities and we neglect the contribution of (\ref{S_pot}).
The time-dependent minimum for $\Psi$ is derived as 
\be
	\langle \Psi \rangle = 
	\bigg(
	   \frac{\kappa^2 |S|^2 + c_2 m_{3/2}^2}{\kappa^2 |S|^2 + c_1 m_{3/2}^2} 
	 \bigg)^{1/4} f_a.
\ee
The equations of motion are given by
\begin{gather}
	\ddot S + 3H \dot S + \kappa^2 (|\Psi|^2 + |\bar\Psi|^2 ) S = 0, \\
	\ddot \Psi + 3H \dot \Psi 
	+ \big[ 
	   \kappa^2 ( |S|^2 + |\bar\Psi|^2 ) + c_1 m_{3/2}^2 
	\big] \Psi 
	- \kappa^2 f_a^2 \bar\Psi^* = 0, \\
	\ddot{\bar\Psi} + 3H \dot{\bar\Psi} 
	+ \big[ \kappa^2 ( |S|^2 + |\Psi|^2 ) + c_2 m_{3/2}^2 
	  \big] \bar\Psi - \kappa^2 f_a^2 \Psi^* = 0.
\end{gather}
Let us redefine the PQ field along the flat direction as $\sigma = 2 [\mathrm{Re}(\Psi) - \langle \Psi \rangle ]$ and call it saxion.
Using this definition, well after the PQ fields approach the minimum, the potential of the saxion is written as $V = 4(\kappa^2 |S|^2 + c_1 m_{3/2}^2) \sigma^2$, so the equation of motion of the saxion becomes
\be
	\ddot{\sigma} + 3H\dot{\sigma} + 8(\kappa^2 |S|^2 + c_1 m_{3/2}^2 ) \sigma 
	+ 2 \langle \ddot{\Psi} \rangle + 6 H \langle \dot{\Psi} \rangle = 0.
\ee
The behavior of the solution of this equation of motion is similar to that of the toy model discussed in the previous section.
While $\kappa S \gg m_{3/2}$, the amplitude of the saxion decays like $t^{-1/2}$ as long as the Hubble parameter does not exceed the effective mass of the PQ fields.
After the saxion mass ($m_{3/2}$) becomes significant ($m_{3/2} \gg \kappa S$), the saxion amplitude decreases as $ t^{-1}$.
The numerical calculations are shown in Fig.~\ref{Fig5}, where time evolutions of scalar fields are plotted.
Similarly to the toy model, the saxion follows its minimum adiabatically even after the temporal minimum shifts to the true minimum at $\kappa S \sim m_{3/2}$.
Thus the resulting amplitude of the saxion is significantly suppressed compared with the ordinary case.

Let us briefly summarize necessary conditions for the adiabatic suppression to work.
\begin{itemize}
\item $\kappa f_a \gg H_{\rm PQ}$ : Otherwise, the effective mass for saxion is smaller than the Hubble parameter and the adiabatic suppression does not work.
\item $\kappa f_a \ll 100H_{\rm PQ}$ : Otherwise, the resonant amplification of saxion is efficient.
Combined with the above condition, the suppression is most efficient for $\kappa f_a \sim 10H_{\rm PQ}$.
\item $\Gamma_S < H_{\rm adi}\equiv H_{\rm PQ}m_{3/2}/(\kappa f_a)$ : 
Otherwise, $S$ decays and large effective mass for the saxion disappears before the saxion relaxes to the true minimum.
\end{itemize}

In the next subsection, we calculate the abundance of the saxion and see how much the saxion coherent oscillation survives.

\subsection{The abundance of the saxion}

As shown in the previous work~\cite{Kawasaki:2007mk}, the abundance of the saxion is strongly constrained from cosmological observations.
If we succeed in reducing the abundance of the saxion, the cosmologically viable model parameter space can be extended, and it is very important when one tries to build a SUSY axion model.
Now, using the setup we developed in the previous subsections, we calculate the abundance of the saxion and see that the adiabatic suppression significantly reduces the final abundance. 

The abundance of the saxion depends on the decay rate of $S$ and decay rate of the inflaton.
Here we assume that the decay of $S$ and inflaton occurs after the saxion relaxes to the true minimum, i.e., $\Gamma_S, \Gamma_I < H_{\rm adi}=H_{\rm PQ}m_{3/2}/(\kappa f_a)$.
In particular, the former condition is necessary for the adiabatic suppression works successfully.
These conditions are written as
\be
	m_{3/2} > \frac{\kappa f_a}{H_\mathrm{PQ}} \Gamma_S ~~~\text{and}~~~ 
	m_{3/2}> \frac{\kappa f_a}{H_\mathrm{PQ}} \Gamma_I.
	\label{suppress_condition}
\ee
In the former condition, the decay rate of $S$ is determined by the interaction of $S$ with the axino. 
Constraints from the axino are discussed in the next subsection.
The latter condition in (\ref{suppress_condition}) is rewritten as 
\be
	m_{3/2} \gtrsim 10^{-6}\,\mathrm{GeV}~
	\bigg( \frac{\kappa f_a}{100H_\mathrm{PQ}} \bigg) 
	\bigg( \frac{T_R}{10^{5}\,\mathrm{GeV}} \bigg)^2.   
	\label{m_sigma}
\ee

Under these assumptions, the energy-to-entropy ratio of the saxion is fixed at the reheating ($H \simeq \Gamma_I$).
By noting that the comoving saxion number density after the resonance ends is adiabatically invariant, the number density is calculated from $n_\sigma(H) = n_\sigma(H_{\rm PQ})[ a(H_{\rm PQ})/a(H) ]^3$ for $H<H_{\rm PQ}$, where $n_\sigma (H_{\rm PQ})=\kappa f_a \sigma_i^2/2=\kappa f_a (\xi f_a)^2/2$.
The energy-to-entropy ratio is then estimated as
\be
	\frac{\rho_\sigma}{s} 
	= \frac{1}{8} T_R \gamma \left( \frac{f_a}{M_P} \right)^2
	\simeq 2.1 \times 10^{-9}\,\mathrm{GeV}\, 
	\gamma \bigg( \frac{T_R}{10^{5}\,\mathrm{GeV}} \bigg) 
	\bigg( \frac{f_a \Delta}{10^{12}\,\mathrm{GeV}} \bigg)^2, 
	\label{abundance}
\ee
where $\gamma$ is a suppression factor defined by 
\be
	\gamma = \xi^2 \bigg( \frac{\kappa f_a}{H_\mathrm{PQ}} \bigg)^2
	\frac{m_{3/2}}{\kappa f_a}
	= 10^{-7} \xi^2 \bigg(\frac{\kappa f_a}{100 H_\mathrm{PQ}} \bigg)^2
	\bigg( \frac{0.1}{\kappa} \bigg)
	\bigg( \frac{10^{12}~\mathrm{GeV}}{f_a} \bigg) 
	\bigg( \frac{m_{3/2}}{1~\mathrm{GeV}} \bigg).
\ee
Here $\Delta$ represents the  difference between the high-energy and low-energy minima, which is written as $\Delta \simeq |1-(c_2/c_1)^{1/4}|$, and this may be order unity in general.
If there was no adiabatic suppression and the parametric resonance, we would have $\gamma = 1$.
Therefore significant suppression is certainly realized unless $\kappa f_a$ is too large compared with the Hubble parameter at the PQ breaking.
Note that the most effective suppression takes place if $\kappa f_a / H_\mathrm{PQ} \sim 10$ because the parametric resonance ends soon and the saxion amplitude is not amplified much.

\begin{figure}[t]
\centering
\subfigure[]{
	\includegraphics [width = 7.5cm, clip]{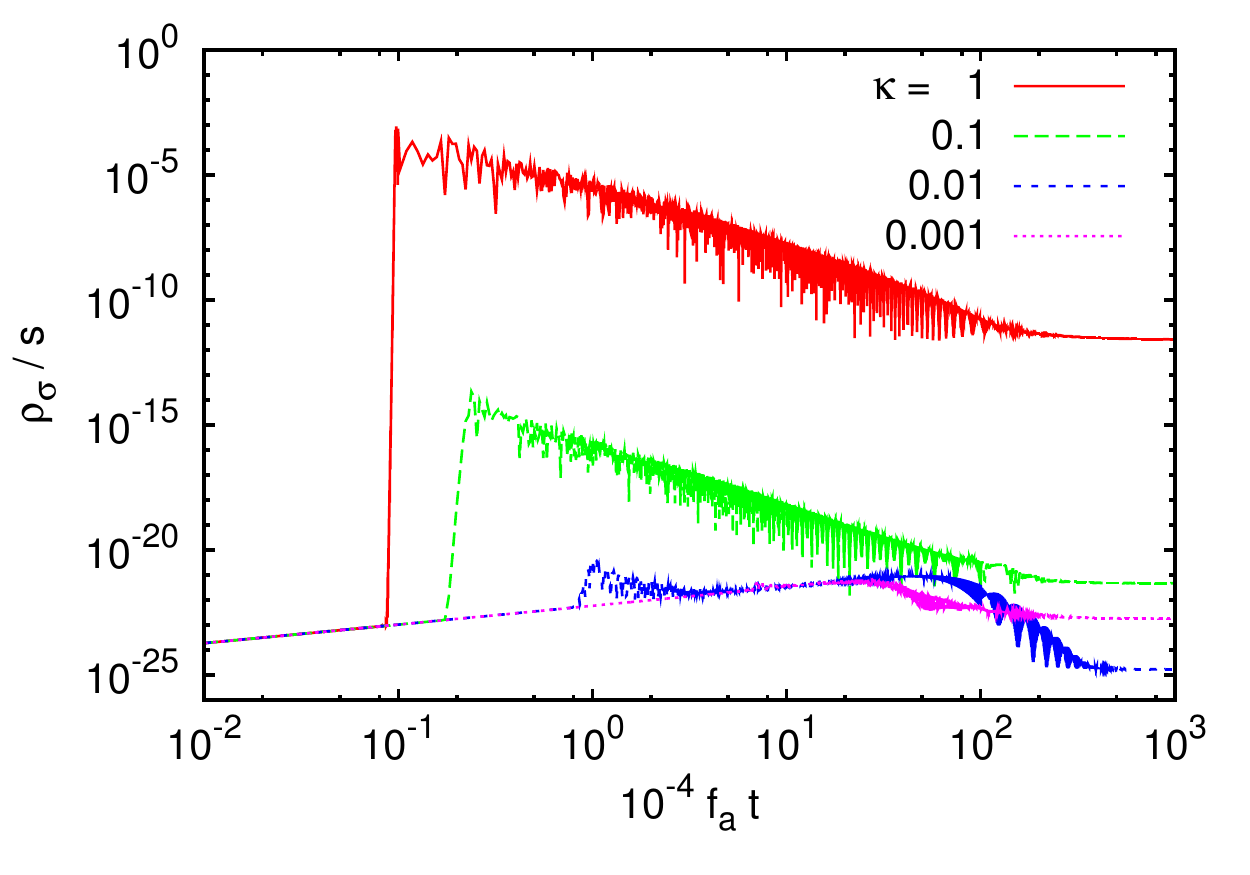}
	\label{Fig7a}
}
\subfigure[]{
	\includegraphics [width = 7.5cm, clip]{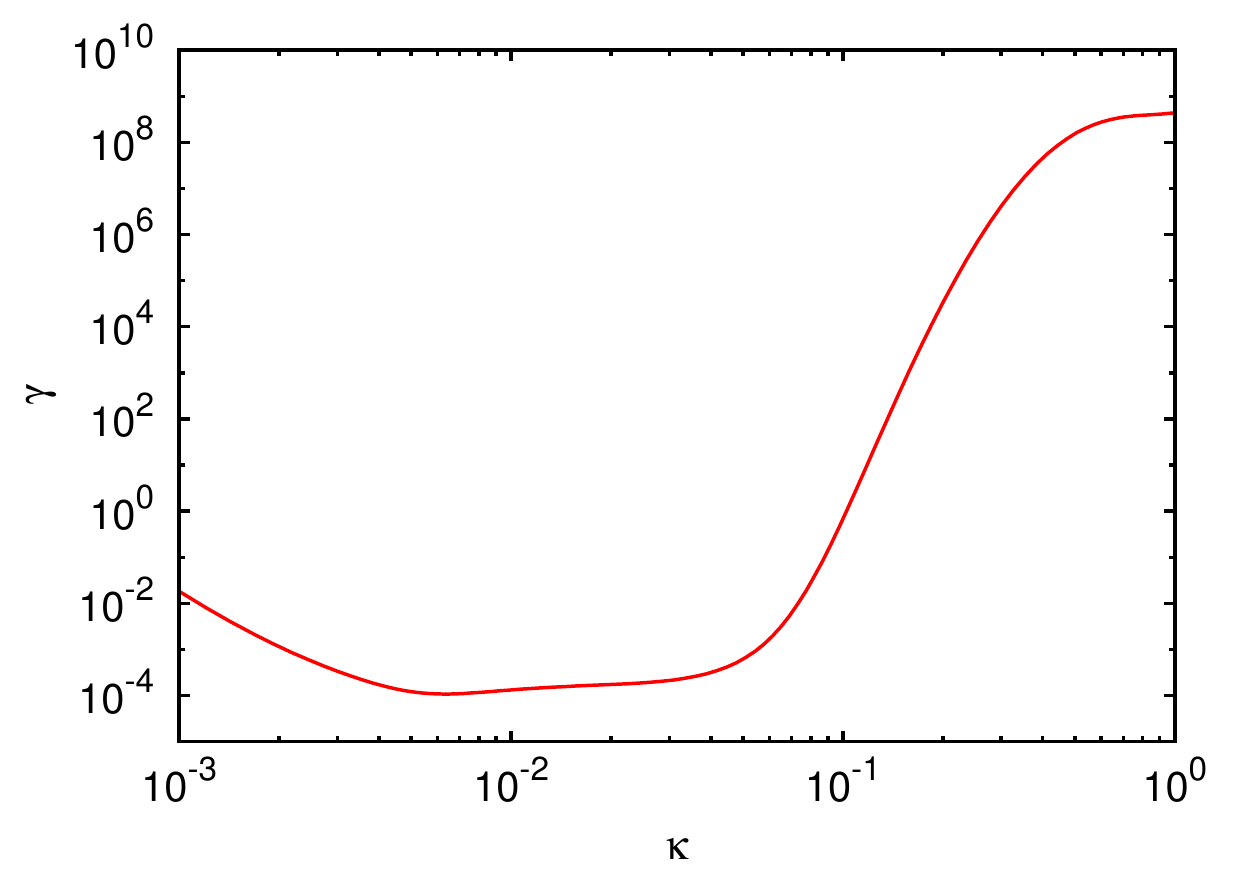}
	\label{Fig7b}
}
\caption{
Time evolutions of $\rho_\sigma / s$ and $\kappa$ dependence of the suppression factor $\gamma$ are shown.
Mass dimensional values are normalized by $f_a$.
We have taken $f_a=10^{12}~\mathrm{GeV}$, $m_{3/2}=10^{-5}$, $\Gamma_S = \Gamma_I = 10^{-6}$, $\Delta=10^{-5}$, $n=4$, $c_H=1$, and $M=10^3$ in both figures.  
In the left figure, we have taken $\kappa = 1$ (solid red line), $\kappa = 0.1$ (dashed green line), $\kappa = 0.01$ (dotted blue line), $\kappa=0.001$ (small-dotted magenta line).
}
\label{Fig7}
\end{figure}

We calculated numerically the evolutions of the energy-to-entropy ratio and $\kappa$-dependence of the suppression factor $\gamma$, and the results are shown in Figure \ref{Fig7}.
In this calculation, mass dimensional values are normalized by $f_a$ and 
in order make numerical calculation easy we have taken $f_a=10^{12}~\mathrm{GeV}$, $m_{3/2}=10^{-5}$, $\Gamma_S=\Gamma_I=10^{-6}$,  $\Delta = 10^{-5}$, $n=4$, $c_H=1$, and $M=10^3$ in Figure \ref{Fig7}.
We found that $\kappa f_a / H_\mathrm{PQ} \sim 10^3,~10^2,~10,~1$ for $\kappa = 1, ~0.1, ~0.01, ~0.001$ respectively in this calculation.
We can see that the most efficient suppression takes place for $\kappa \sim 0.01$ and the resonant amplification becomes efficient for $\kappa > 0.1$.
For $\kappa < 0.001$, the adiabatic suppression no longer works and $\gamma$ approaches unity.
Note that $\gamma$ becomes much smaller than unity in realistic models even if the resonant amplification takes place most efficiently, because the realistic gravitino mass is many orders of magnitude smaller than that we use in our numerical calculation.

\begin{figure}[t]
\centering
\subfigure[$f_a = 10^{10}~\mathrm{GeV}$]{
	\includegraphics [width = 7.5cm, clip]{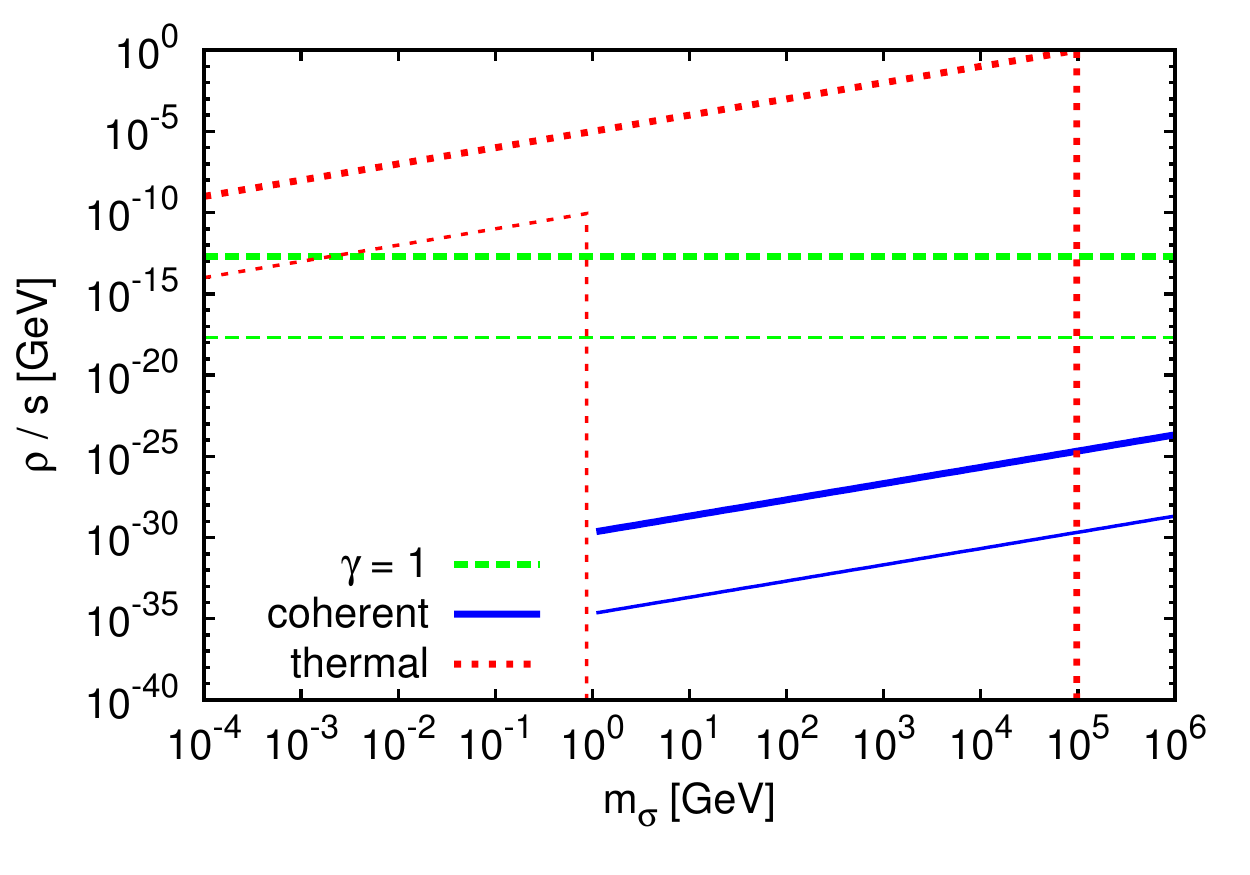}
	\label{Fig6a}
}
\subfigure[$f_a = 10^{12}~\mathrm{GeV}$]{
	\includegraphics [width = 7.5cm, clip]{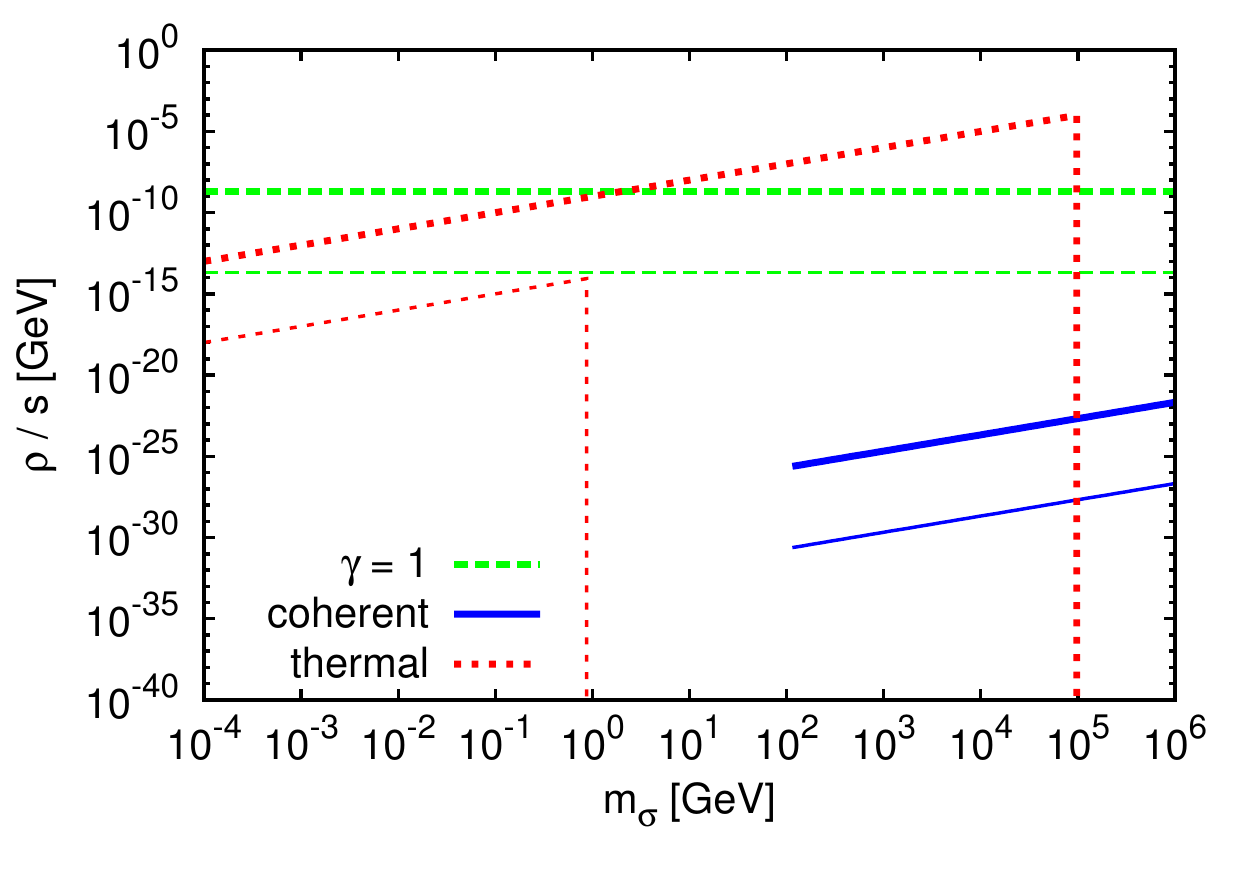}
	\label{Fig6b}
}
\caption{
The mass dependence of the energy-to-entropy ratio of the saxion predicted by our analysis.
The solid blue lines represent the our results for the saxion coherent oscillation, the dashed green lines represent ordinary analysis of the saxion coherent oscillation without the adiabatic suppression ($\gamma = 1)$, and the red dotted lines represent the thermally-produced saxion.
Thick and thin lines correspond to the reheating temperature $T_R=10^5~\mathrm{GeV}$ and $T_R=1~\mathrm{GeV}$ respectively.
In this calculation, we have taken $\kappa f_a / H_\mathrm{PQ} =10$, $\kappa=0.001$, $\xi=10^{-6}$ for both figures and $f_a=10^{10}~(10^{12})~\mathrm{GeV}$ for left (right) figure.
The solid blue lines disappear at the point where the adiabatic suppression (\ref{suppress_condition}), which is rewritten as (\ref{cond-m_3/2}), is broken.
}
\label{Fig6}
\end{figure}

The saxion is also produced from thermal bath and the contribution to the saxion abundance should not be neglected.
The temperature at which the saxion decouples from the thermal bath is estimated as~\cite{Rajagopal:1990yx}
\be
	T_D \sim 10^{11}\,\mathrm{GeV}
	\bigg( \frac{f_a}{10^{12}\,\mathrm{GeV}} \bigg)^2.
\ee
The abundance of the thermal saxion is then estimated as
\be
	\bigg( \frac{\rho_\sigma}{s} \bigg)^{(\text{TP})} 
	\sim 10^{-3}\,\mathrm{GeV}\,\bigg( \frac{m_{3/2}}{1\,\mathrm{GeV}} \bigg)
	 ~~\text{for}~~T_R > T_D,
\ee
and
\be
	\bigg( \frac{\rho_\sigma}{s} \bigg)^{(\text{TP})} 
	\sim 10^{-9}\,\mathrm{GeV}\,\bigg( \frac{m_{3/2}}{1\,\mathrm{GeV}} \bigg)
	\bigg( \frac{T_R}{10^5\,\mathrm{GeV}} \bigg) 
	\bigg( \frac{10^{12}\,\mathrm{GeV}}{f_a} \bigg)^2 
	~~\text{for}~~T_R < T_D.
\ee
These predictions are summarized in Fig.~\ref{Fig6}.
Our main result of the adiabatically suppressed saxion is shown as the solid blue lines 
and the dashed green lines represent the saxion coherent oscillation without the adiabatic suppression ($\gamma = 1)$ and the red dotted lines represent the thermally-produced saxion.
The solid blue lines disappear at the point where the adiabatic suppression (\ref{suppress_condition}), which is rewritten as (\ref{cond-m_3/2}), is broken.It is shown that, considering the case of the small $\xi$, in which $\kappa f_a / H_{\mathrm{PQ}} \sim 10$ and $\kappa = 0.001$, the abundance of the coherent oscillation becomes much smaller and completely negligible compared to the thermal saxion.
In particular, it is remarkable that the cosmological constraints for the saxion is much weakened than previously thought~\cite{Kawasaki:2007mk} at the region $m_\sigma > T_R$.

Note that since the PQ symmetry is broken after inflation, cosmic strings are formed at the PQ phase transition.
Moreover, at the QCD phase transition, domain walls appear~\cite{Sikivie:1982qv}.
In order to avoid the cosmological domain wall problem, the color anomaly number must be equal to one.
In this case domain walls are bounded by strings, and walls as well as strings disappear due to the wall tension~\cite{Vilenkin:1982ks,Vachaspati:1984dz,Nagasawa:1994qu,Chang:1998tb,Hiramatsu:2010yn}.
The PQ scale, $f_a$, cannot be larger than $\sim 10^{12}$\,GeV in this case since axions produced through coherent oscillation and emission from the strings and walls contribute to the dark matter density~\cite{Yamaguchi:1998gx,Hiramatsu:2010yu}.

\subsection{Constraints from the axino}

The abundance of the axino is also strongly constrained from the cosmological observations.
In the present model, the axinos are produced non-thermally through the interaction with $S$.
From the superpotential (\ref{superpot}), the interaction Lagrangian is given by
\be
	\mathcal{L}_{S \text{-} \tilde a} = -\frac{1}{2}\kappa S \tilde a \tilde a - {\rm h.c.},
	\label{S-axino_int}
\ee
where the axino $\tilde a$ is defined using the fermionic component of the PQ fields, $\psi$ and $\bar\psi$, as $\tilde a = (\psi - \bar\psi) /\sqrt 2$.
The decay rate of $S$ into the axino pair is calculated as 
\be
	\Gamma_{S \to \tilde a \tilde a} = \frac{\sqrt 2}{32 \pi} \kappa^3 f_a.
\ee
In order for the adiabatic suppression to work, it is necessary to satisfy the condition (\ref{suppress_condition}).
The condition is rewritten as 
\be
	m_{3/2} > 1~ \mathrm{TeV} \bigg( \frac{\kappa}{10^{-2}} \bigg)^3 \bigg( \frac{f_a}{10^{10}~\mathrm{GeV}} \bigg) \bigg( \frac{\kappa f_a / H_\mathrm{PQ}}{10} \bigg).
	\label{cond-m_3/2}
\ee
Hence the adiabatic suppression works for the gravitino mass of order of TeV scale.
The lighter gravitino mass is possible if we choose $\kappa \lesssim 10^{-3}$.
In such a case, the efficient adiabatic suppression can be realized for larger $M$ to satisfy $\kappa f_a \simeq 10H_{\rm PQ}$.
The parameter set chosen in Fig.~\ref{Fig4c} ($\kappa=10^{-3}$, $M=5 \times 10^{3}$ and $H=6 \times 10^{-4}$) satisfies the condition (\ref{cond-m_3/2}) and we have found $\xi \sim 10^{-6}$.

Next, we calculate the axino abundance and derive cosmological constraints on the parameters.
Before calculating the abundance, we comment on the axino mass.
The axino mass is given by the Yukawa interaction (\ref{S-axino_int}).
Due to the supergravity effect, the expectation value of $\kappa S$ is not zero but of the order of the gravitino mass.
Thus we denote the axino mass by $m_{\tilde a} = \kappa \langle S \rangle \sim m_{3/2}$.
Assuming that the decay of $S$ occurs before the reheating and the other decay modes except for that into the axino are negligible, 
the axino abundance from the decay of $S$ is calculated as
\be
	Y_{\tilde a}^\mathrm{(NTP)} 
	= \frac{1}{8} \frac{T_R}{m_{\tilde a}} \frac{\kappa^2 f_a^4}{M_P^2 H_\mathrm{PQ}^2}
	\simeq 2 \times 10^{-12} \bigg( \frac{T_R}{m_{\tilde a}} \bigg) \bigg( \frac{f_a}{10^{12}~\mathrm{GeV}} \bigg)^2 \bigg( \frac{\kappa f_a / H_\mathrm{PQ}}{10} \bigg)^2.
\ee
On the other hand, the observational upper limit is given as $Y_{\tilde a} \lesssim 10^{-12}$ if the axino decays into the LSPs 
and $m_{\tilde a} Y_{\tilde a} \lesssim 4 \times 10^{-10}~\mathrm{GeV}$ if the axino is the LSP.
These constraints are satisfied in the present model if $T_R \lesssim 1~\mathrm{TeV}$ and $m_{\tilde a} \sim 1~\mathrm{TeV}$.

The axinos produced from the thermal bath after reheating also contributes the resultant abundance \cite{Covi:2001nw,Brandenburg:2004du} and it is estimated as
\be
	Y^\mathrm{(TP)}_{\tilde a} \simeq 2.0 \times 10^{-10} g_s^6 \ln \bigg( \frac{1.108}{g_s} \bigg) \bigg( \frac{10^{12}~\mathrm{GeV}}{f_a} \bigg)^2 \bigg( \frac{T_R}{1~\mathrm{TeV}} \bigg),
	\label{axino_abun_TP}
\ee
whare $g_s$ is a strong coupling constant which is order unity.
To avoid this constraint, we need the relatively light axino with high $f_a$ ($m_{\tilde a} \lesssim 1~\mathrm{GeV}$ for $f_a \sim 10^{12}~\mathrm{GeV}$), or relatively heavy axino whose mass exceeds the reheating temperature ($m_{\tilde a} \gtrsim T_R$), since such axinos cannot be produced in a thermal bath and the estimate (\ref{axino_abun_TP}) is no longer applied.
As a summery of this subsection, we conclude that the adiabatic suppression mechanism can be applied to the SUSY axion model without inducing the cosmological axino problem.

\section{Conclusions} \label{conc}

We revisited the problems induced by the scalar coherent oscillation such as the moduli or saxion.
In the ordinary consideration, the scalar field initially displaced far from the origin begins to oscillate when the Hubble parameter becomes equal to the mass of the scalar field, which leads to the large abundance and often becomes harmful for cosmology.

In this paper, we have proposed a model in which the amplitude of oscillating scalar fields is significantly suppressed by using the idea of adiabatic suppression.
This mechanism works if the effective mass of the scalar field is much lager than the Hubble parameter.
We found that such a situation is realized by coupling the scalar field with another scalar field, whose initial amplitude is also far displaced from the origin.
We have investigated these scaler field dynamics and seen that the oscillating scalar field follows the temporal minimum of the potential adiabatically without inducing a large oscillation amplitude depending on model parameters.
In the model we have analyzed, we have encountered a parametric resonance for amplifying the oscillation amplitude, but still the abundance can be significantly reduced.
Although the resulting abundance depends on the model parameters, a broad parameter region is available for successful adiabatic suppression.

We have applied the adiabatic suppression mechanism to the SUSY axion model.
We followed the dynamics of the PQ scalar fields and calculated the resulting abundance of the saxion.
According to the result, the cosmological constraint for the saxion is greatly relaxed particularly for large PQ breaking scale 
without relying on any additional entropy production.
Although we have focused on the SUSY axion model as a concrete example, we believe that the idea has a broad applicability.
A similar mechanism may be used for solving the cosmological problems associated with scalar field oscillation appearing in many models beyond SM.

\section*{Acknowledgment}

This work is supported by Grant-in-Aid for Scientific research from
the Ministry of Education, Science, Sports, and Culture (MEXT), Japan,
No.\ 14102004 (M.K.), No.\ 21111006 (M.K. and K.N.), No.\ 22244030 (K.N.) and also 
by World Premier International Research Center
Initiative (WPI Initiative), MEXT, Japan.

  

\end{document}